\documentclass[acmtoms]{acmsmall}
\usepackage{hyperref}
\usepackage{amsmath,amsfonts,amssymb}
\usepackage{xspace}
\usepackage{geometry}
\usepackage{pgf,tikz}
\usetikzlibrary{trees,quadtree,thplex,matrix,arrows,patterns,3d}

\usepackage[capitalize]{cleveref}
\usepackage[ruled,noline,linesnumbered]{algorithm2e}

\tikzset{subref/.style={anchor=north west,execute at begin node={\textcolor{black}{(}}, execute at end node={)}}}
                 
\DeclareMathOperator{\Star}{star}
\DeclareMathOperator{\supp}{supp}
\DeclareMathOperator{\cone}{cone}
\DeclareMathOperator{\clos}{clos}

\DeclareMathOperator{\parent}{parent}
\DeclareMathOperator{\children}{children}
\newcommand{\tr}{\mathsf{T}}
\newcommand{\half}{\textstyle\frac12}
\newcommand{\Curl}{\text{curl}}
\newcommand{\Div}{\text{div}}
\newcommand{\Nabla}{\boldsymbol{\nabla}}

\SetCommentSty{mycomment}
\SetArgSty{texttt}
\SetKwIF{If}{ElseIf}{Else}{if}{\{}{else if}{else}{\}}
\SetKwFor{For}{for}{\{}{\}}
\SetKwSty{texttt}
\crefname{algocf}{Ex.}{Exs.}
\Crefname{algocf}{Example}{Examples}

\newcommand{\ierr}{}
\newcommand{\chkerr}{}
\newcommand{\fn}[1]{\textcolor{green!50!black}{#1}}
\newcommand{\nll}{\textcolor{gray}{NULL}}

\makeatletter
\def\pd{\@ifnextchar[{\@pdwith}{\@pdwithout}}
\def\@pdwith[#1]#2#3{\href{http://www.mcs.anl.gov/petsc/petsc-3.6/docs/manualpages/#2/#1.html}{#3}}
\def\@pdwithout#1#2{\@pdwith[#2]{#1}{#2}}
\makeatother

\begin{document}

\markboth{T.~Isaac and M.~G.~Knepley}{%
  Support for Non-conformal Meshes in PETSc's DMPlex Interface%
}%

\title{Support for Non-conformal Meshes in PETSc's DMPlex Interface}

\author{%
  TOBIN ISAAC \affil{The University of Chicago}
  MATTHEW G.~KNEPLEY \affil{Rice University}
}%

\begin{abstract}
  PETSc's DMPlex interface for unstructured meshes has been extended to
  support non-conformal meshes.  The topological construct that DMPlex
  implements---the CW-complex---is by definition conformal, so representing
  non-conformal meshes in a way that hides complexity requires careful
  attention to the interface between DMPlex and numerical methods such as the
  finite element method.  Our approach---which combines a tree structure for
  subset-superset relationships and a ``reference tree'' describing the types
  of non-conformal interfaces---allows finite element code written for conformal
  meshes to extend automatically: in particular, all ``hanging-node''
  constraint calculations are handled behind the scenes.  We give example code
  demonstrating the use of this extension, and use it to convert forests of
  quadtrees and forests of octrees from the p4est library to DMPlex meshes.
\end{abstract}


\acmformat{%
  Tobin Isaac and Matthew G.~Knepley, 2015. Support for Non-conformal Meshes
  in PETSc's DMPlex Interface.%
}%

\begin{bottomstuff}
\end{bottomstuff}

\maketitle

\section{Introduction}

PETSc \cite{petsc-user-ref,petsc-web-page} is an actively developed and widely
used library for numerical methods in scientific computing, providing parallel
data management, structured and unstructured meshes, linear and nonlinear
algebraic solvers and preconditioners, time integrators and optimization
algorithms.  Many of these methods (such as geometric multigrid and domain
decomposition linear and nonlinear solvers) can take advantage of the
geometric/topological setting of a discretized problem, i.e. mesh information.
PETSc's interface for serving mesh data to numerical algorithms is the
{\bf\pd{DM}{DM}} object.  PETSc has native DM implementations for several mesh
formats, and implementations that wrap external libraries may also be
registered, such as {\bf\pd{DM}{DMMOAB}} for MOAB \cite{tautges_moab:_2004}.
Because of PETSc's pointer-to-implementation approach to method extensibility,
external implementations may cover only those methods in the DM API that are
necessary for their target applications.  While no DM implementation is
privileged above others---PETSc-native and external implementations are
registered in the same way---the native implementations of structured grids
({\bf\pd{DM}{DMDA}}) and unstructured meshes ({\bf\pd[DMPLEX]{DM}{DMPlex}})
have the most complete coverage of the DM API, and are developed most actively.
Only DMPlex, for example, currently has complete support for the
{\bf\pd{DM}{PetscFE}} and {\bf\pd{DM}{PetscFV}} implementations of the finite
element method and finite volume method.

Many mesh formats lie between structured grids and unstructured meshes and can
broadly be described as hierarchical mesh formats.  Examples include red-green
refinement of triangular meshes, quadtree/octree refinement of
quadrilateral/hexahedral meshes, and nested Cartesian grids.  These formats
are often implemented in frameworks with data structures that are advantageous
for certain data access patterns.  Patch-based nested grids, for instance,
are optimized for fast stencil operations.  Another example is red-green
refined triangular meshes, where the triangles have been ordered by a
Sierpinski curve: these meshes can efficiently compute residuals of
discontinuous-Galerkin or finite volume operators without explicitly
determining which cells are adjacent to each-other by pushing fluxes onto
stacks \cite{bader2012memory}.

We would like to be able to convert hierarchical mesh formats into the DMPlex
unstructured format.  Our main reason is that the efficiency gains of
hierarchical mesh formats typically come at the expense of flexibility and
generality.  DMPlex supports, for instance, arbitrary mesh partitions and the
extraction of arbitrary subsets of cells (or facets) as submeshes: features
which are typically missing from hierarchical meshing frameworks.  The broad
support of DMPlex for PETSc's DM API also makes it an ideal format for testing
and comparing numerical methods that call on the DM interface.

What has prevented the conversion of these meshes in the past is that
hierarchical meshes are often non-conformal meshes: this is true of quadtrees,
octrees, nested Cartesian grids, and of some hierarchical simplicial meshes as
well.  CW-complexes---the topologies DMPlex was designed to represent---are by
definition conformal.  Our recent extension of DMPlex has addressed this
shortcoming.  In this paper we describe how we represent non-conformal
hierarchical meshes in DMPlex in a way that minimally disturbs the way DMPlex
interacts with the other components of PETSc, and that requires minimal input
from the user.

\section{Preliminaries: conformal meshes}
\label{sec:conformal}

We begin with a brief review of the DMPlex interface and the finite
element method in the context of conformal meshes.

\subsection{CW-complexes and DMPlex}

The triangulation of a domain $\Omega$ into cells generates a CW-complex (see,
e.g., \cite[Chapter 10]{hatcheralgebraic}).  In short, a CW-complex is a
partition of a $d$-dimensional spaces into well-shaped open cells with
dimensions between $0$ and $d$, such that the boundary of each $n$-cell
($n>0$) is partitioned by finitely many lower-dimensional cells.  We call
cells of every dimension ``points'' in the complex.

In a CW-complex, the basic relationship that defines the topology is the map
from an $n$-cell $A$ to the $(n-1)$-cells on its boundary, which we call the
cone of $A$, $\cone(A)$, following the terminology in \cite{knepley2009mesh}.
The closure of the cone map,
\begin{equation}\label{eq:clos}
  \clos(A) := \{A\} \cup \cone(A) \cup \cone(\cone(A)) \cup \dots,
\end{equation}
corresponds to closure in $\overline{\Omega}$, i.e., $\clos(A)$ partitions
$\overline{A}$.

The reverse map, taking the $n$-cell $A$ to its adjacent $(n+1)$-cells, is
called the support map, $\supp(A)$, and the closure of the support map is
called the star of $A$, $\Star(A)$.  It is important to note at this point
that for conformal meshes, cones and supports are dual,
\begin{equation}
  B \in \clos(A) \Leftrightarrow A \in \supp(B).
\end{equation}

A CW-complex can be represented by a Hasse diagram for stratified
partially-ordered sets: the depth of a stratum corresponds to the topological
dimension of its points, upward arrows represent cone maps, and downward
arrows represent support maps.  These concepts (cones, supports, strata) are
at the core of the DMPlex interface, which we illustrate in
\cref{fig:twotriangle:base,fig:twotriangle:cone,fig:twotriangle:clos,%
fig:twotriangle:supp,fig:twotriangle:star}.

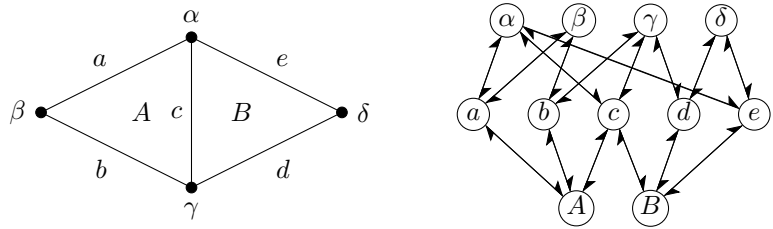
\begin{figure}\centering
  \begin{tikzpicture}
    \begin{scope}[local bounding box=box]
      \node (mesh) {%
        \begin{tikzpicture}
  \usetikzlibrary{thplex}
  \tikzset{Acone/.prefix style={}}
  \tikzset{Aclos/.prefix style={}}
  \tikzset{Aclosfill/.prefix style={}}
  \tikzset{deltasupp/.prefix style={}}
  \tikzset{deltastar/.prefix style={}}
  \tikzset{deltastarfill/.prefix style={}}
  \coordinate (al) at ( 0, 1);
  \coordinate (be) at (-2, 0);
  \coordinate (ga) at ( 0,-1);
  \coordinate (de) at ( 2, 0);
  \fill [cell,Aclos,Aclosfill] (al) -- (be) -- (ga) -- cycle
  node at (barycentric cs:al=1,be=1,ga=1) {$A$};
  \fill [cell,deltastar,deltastarfill] (al) -- (ga) -- (de) -- cycle
  node at (barycentric cs:al=1,ga=1,de=1) {$B$};
  \draw [facet,Acone,Aclos] (al) -- node [above left] {$a$} (be);
  \draw [facet,Acone,Aclos] (be) -- node [below left] {$b$} (ga);
  \draw [facet,Acone,Aclos] (ga) -- node [left] {$c$} (al);
  \draw [facet,deltasupp,deltastar] (ga) -- node [below right] {$d$} (de);
  \draw [facet,deltasupp,deltastar] (de) -- node [above right] {$e$} (al);
  \node at (al) [vertex,Aclos,label={[vertexlabel,Aclos]above:$\alpha$}] {};
  \node at (be) [vertex,Aclos,label={[vertexlabel,Aclos]left:$\beta$}] {};
  \node at (ga) [vertex,Aclos,label={[vertexlabel,Aclos]below:$\gamma$}] {};
  \node at (de) [vertex,deltastar,label={[vertexlabel,deltastar]right:$\delta$}] {};
\end{tikzpicture}
      };%
      \node [right=2cm,right of=mesh,anchor=west] (hasse) {%
        \begin{tikzpicture}
  \usetikzlibrary{thplex}
  \tikzset{Acone/.prefix style={}}
  \tikzset{Aclos/.prefix style={}}
  \tikzset{deltasupp/.prefix style={}}
  \tikzset{deltastar/.prefix style={}}
  \matrix (cells) {%
    \node [point,Aclos] (hA) {$A$};
    &
    \node [point,deltastar] (hB) {$B$};
    \\
    };%
    \matrix [above of=cells] (facets) {%
      \node [point,Acone,Aclos] (ha) {$a$};
      &
      \node [point,Acone,Aclos] (hb) {$b$};
      &
      \node [point,Acone,Aclos] (hc) {$c$};
      &
      \node [point,deltasupp,deltastar] (hd) {$d$};
      &
      \node [point,deltasupp,deltastar] (he) {$e$};
      \\
      };%
      \matrix [above of=facets] (vertices) {%
        \node [point,Aclos] (hal) {$\alpha$};
        &
        \node [point,Aclos] (hbe) {$\beta$};
        &
        \node [point,Aclos] (hga) {$\gamma$};
        &
        \node [point,deltastar] (hde) {$\delta$};
        \\
        };%
        \draw[arrow] (he) -- (hal);
        \draw[arrow] (hal) -- (he);
        \draw[arrow]             (ha) -- (hA);
        \draw[arrow,Acone,Aclos] (hA) -- (ha);
        \draw[arrow]             (hb) -- (hA);
        \draw[arrow,Acone,Aclos] (hA) -- (hb);
        \draw[arrow]             (hc) -- (hA);
        \draw[arrow,Acone,Aclos] (hA) -- (hc);
        \draw[arrow] (hB) -- (hc);
        \draw[arrow] (hc) -- (hB);
        \draw[arrow]           (hB) -- (hd);
        \draw[arrow,deltastar] (hd) -- (hB);
        \draw[arrow]           (hB) -- (he);
        \draw[arrow,deltastar] (he) -- (hB);
        \draw[arrow]       (hal) -- (ha);
        \draw[arrow,Aclos] (ha) -- (hal);
        \draw[arrow]       (hbe) -- (ha);
        \draw[arrow,Aclos] (ha) -- (hbe);
        \draw[arrow]       (hbe) -- (hb);
        \draw[arrow,Aclos] (hb) -- (hbe);
        \draw[arrow]       (hga) -- (hb);
        \draw[arrow]       (hga) -- (hc);
        \draw[arrow,Aclos] (hb) -- (hga);
        \draw[arrow]       (hal) -- (hc);
        \draw[arrow,Aclos] (hc) -- (hga);
        \draw[arrow] (hd) -- (hga);
        \draw[arrow,Aclos] (hc) -- (hal);
        \draw[arrow] (hga) -- (hd);
        \draw[arrow]                     (hd) -- (hde);
        \draw[arrow,deltasupp,deltastar] (hde) -- (hd);
        \draw[arrow]                     (he) -- (hde);
        \draw[arrow,deltasupp,deltastar] (hde) -- (he);
      \end{tikzpicture}

      };%
    \end{scope}
  \end{tikzpicture}
  \caption{A two-triangle mesh and its Hasse diagram.}%
  \label{fig:twotriangle:base}%
\end{figure}

{%
  \tikzset{point/.append style={gray}}
  \tikzset{vertex/.append style={gray}}
  \tikzset{vertexlabel/.append style={gray}}
  \tikzset{facet/.append style={gray}}
  \tikzset{cell/.append style={gray}}
  \tikzset{every picture/.append style={draw=gray,fill=gray}}

  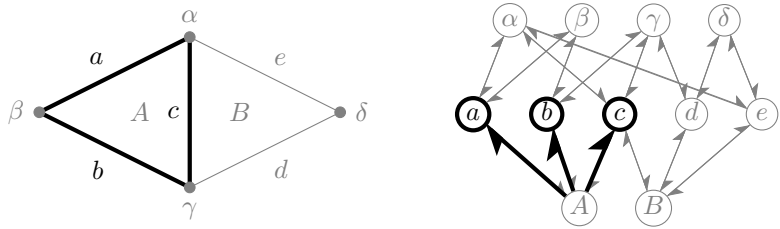
\begin{figure}\centering
    \begin{tikzpicture}
      \begin{scope}[local bounding box=box]
        \tikzset{Acone/.append style={black,ultra thick}}
        \node (Acone) {%
          \begin{tikzpicture}
  \usetikzlibrary{thplex}
  \tikzset{Acone/.prefix style={}}
  \tikzset{Aclos/.prefix style={}}
  \tikzset{Aclosfill/.prefix style={}}
  \tikzset{deltasupp/.prefix style={}}
  \tikzset{deltastar/.prefix style={}}
  \tikzset{deltastarfill/.prefix style={}}
  \coordinate (al) at ( 0, 1);
  \coordinate (be) at (-2, 0);
  \coordinate (ga) at ( 0,-1);
  \coordinate (de) at ( 2, 0);
  \fill [cell,Aclos,Aclosfill] (al) -- (be) -- (ga) -- cycle
  node at (barycentric cs:al=1,be=1,ga=1) {$A$};
  \fill [cell,deltastar,deltastarfill] (al) -- (ga) -- (de) -- cycle
  node at (barycentric cs:al=1,ga=1,de=1) {$B$};
  \draw [facet,Acone,Aclos] (al) -- node [above left] {$a$} (be);
  \draw [facet,Acone,Aclos] (be) -- node [below left] {$b$} (ga);
  \draw [facet,Acone,Aclos] (ga) -- node [left] {$c$} (al);
  \draw [facet,deltasupp,deltastar] (ga) -- node [below right] {$d$} (de);
  \draw [facet,deltasupp,deltastar] (de) -- node [above right] {$e$} (al);
  \node at (al) [vertex,Aclos,label={[vertexlabel,Aclos]above:$\alpha$}] {};
  \node at (be) [vertex,Aclos,label={[vertexlabel,Aclos]left:$\beta$}] {};
  \node at (ga) [vertex,Aclos,label={[vertexlabel,Aclos]below:$\gamma$}] {};
  \node at (de) [vertex,deltastar,label={[vertexlabel,deltastar]right:$\delta$}] {};
\end{tikzpicture}
        };%
        \node [right=2cm,right of=Acone,anchor=west] (Aconehasse) {%
          \begin{tikzpicture}
  \usetikzlibrary{thplex}
  \tikzset{Acone/.prefix style={}}
  \tikzset{Aclos/.prefix style={}}
  \tikzset{deltasupp/.prefix style={}}
  \tikzset{deltastar/.prefix style={}}
  \matrix (cells) {%
    \node [point,Aclos] (hA) {$A$};
    &
    \node [point,deltastar] (hB) {$B$};
    \\
    };%
    \matrix [above of=cells] (facets) {%
      \node [point,Acone,Aclos] (ha) {$a$};
      &
      \node [point,Acone,Aclos] (hb) {$b$};
      &
      \node [point,Acone,Aclos] (hc) {$c$};
      &
      \node [point,deltasupp,deltastar] (hd) {$d$};
      &
      \node [point,deltasupp,deltastar] (he) {$e$};
      \\
      };%
      \matrix [above of=facets] (vertices) {%
        \node [point,Aclos] (hal) {$\alpha$};
        &
        \node [point,Aclos] (hbe) {$\beta$};
        &
        \node [point,Aclos] (hga) {$\gamma$};
        &
        \node [point,deltastar] (hde) {$\delta$};
        \\
        };%
        \draw[arrow] (he) -- (hal);
        \draw[arrow] (hal) -- (he);
        \draw[arrow]             (ha) -- (hA);
        \draw[arrow,Acone,Aclos] (hA) -- (ha);
        \draw[arrow]             (hb) -- (hA);
        \draw[arrow,Acone,Aclos] (hA) -- (hb);
        \draw[arrow]             (hc) -- (hA);
        \draw[arrow,Acone,Aclos] (hA) -- (hc);
        \draw[arrow] (hB) -- (hc);
        \draw[arrow] (hc) -- (hB);
        \draw[arrow]           (hB) -- (hd);
        \draw[arrow,deltastar] (hd) -- (hB);
        \draw[arrow]           (hB) -- (he);
        \draw[arrow,deltastar] (he) -- (hB);
        \draw[arrow]       (hal) -- (ha);
        \draw[arrow,Aclos] (ha) -- (hal);
        \draw[arrow]       (hbe) -- (ha);
        \draw[arrow,Aclos] (ha) -- (hbe);
        \draw[arrow]       (hbe) -- (hb);
        \draw[arrow,Aclos] (hb) -- (hbe);
        \draw[arrow]       (hga) -- (hb);
        \draw[arrow]       (hga) -- (hc);
        \draw[arrow,Aclos] (hb) -- (hga);
        \draw[arrow]       (hal) -- (hc);
        \draw[arrow,Aclos] (hc) -- (hga);
        \draw[arrow] (hd) -- (hga);
        \draw[arrow,Aclos] (hc) -- (hal);
        \draw[arrow] (hga) -- (hd);
        \draw[arrow]                     (hd) -- (hde);
        \draw[arrow,deltasupp,deltastar] (hde) -- (hd);
        \draw[arrow]                     (he) -- (hde);
        \draw[arrow,deltasupp,deltastar] (hde) -- (he);
      \end{tikzpicture}

        };%
      \end{scope}
    \end{tikzpicture}
    \caption{$\cone(A)$ / {\bf \protect\pd{DM}{DMPlexGetCone}()}.}%
    \label{fig:twotriangle:cone}%
  \end{figure}

  \begin{figure}\centering
      \begin{tikzpicture}
        \begin{scope}[local bounding box=box]
          \tikzset{Aclos/.append style={black,ultra thick}}
          \tikzset{Aclosfill/.append style={fill=gray!25!white}}
          \node (Aclos) {%
            \begin{tikzpicture}
  \usetikzlibrary{thplex}
  \tikzset{Acone/.prefix style={}}
  \tikzset{Aclos/.prefix style={}}
  \tikzset{Aclosfill/.prefix style={}}
  \tikzset{deltasupp/.prefix style={}}
  \tikzset{deltastar/.prefix style={}}
  \tikzset{deltastarfill/.prefix style={}}
  \coordinate (al) at ( 0, 1);
  \coordinate (be) at (-2, 0);
  \coordinate (ga) at ( 0,-1);
  \coordinate (de) at ( 2, 0);
  \fill [cell,Aclos,Aclosfill] (al) -- (be) -- (ga) -- cycle
  node at (barycentric cs:al=1,be=1,ga=1) {$A$};
  \fill [cell,deltastar,deltastarfill] (al) -- (ga) -- (de) -- cycle
  node at (barycentric cs:al=1,ga=1,de=1) {$B$};
  \draw [facet,Acone,Aclos] (al) -- node [above left] {$a$} (be);
  \draw [facet,Acone,Aclos] (be) -- node [below left] {$b$} (ga);
  \draw [facet,Acone,Aclos] (ga) -- node [left] {$c$} (al);
  \draw [facet,deltasupp,deltastar] (ga) -- node [below right] {$d$} (de);
  \draw [facet,deltasupp,deltastar] (de) -- node [above right] {$e$} (al);
  \node at (al) [vertex,Aclos,label={[vertexlabel,Aclos]above:$\alpha$}] {};
  \node at (be) [vertex,Aclos,label={[vertexlabel,Aclos]left:$\beta$}] {};
  \node at (ga) [vertex,Aclos,label={[vertexlabel,Aclos]below:$\gamma$}] {};
  \node at (de) [vertex,deltastar,label={[vertexlabel,deltastar]right:$\delta$}] {};
\end{tikzpicture}
          };%
          \node [right=2cm,right of=Aclos,anchor=west] (Acloshasse) {%
            \begin{tikzpicture}
  \usetikzlibrary{thplex}
  \tikzset{Acone/.prefix style={}}
  \tikzset{Aclos/.prefix style={}}
  \tikzset{deltasupp/.prefix style={}}
  \tikzset{deltastar/.prefix style={}}
  \matrix (cells) {%
    \node [point,Aclos] (hA) {$A$};
    &
    \node [point,deltastar] (hB) {$B$};
    \\
    };%
    \matrix [above of=cells] (facets) {%
      \node [point,Acone,Aclos] (ha) {$a$};
      &
      \node [point,Acone,Aclos] (hb) {$b$};
      &
      \node [point,Acone,Aclos] (hc) {$c$};
      &
      \node [point,deltasupp,deltastar] (hd) {$d$};
      &
      \node [point,deltasupp,deltastar] (he) {$e$};
      \\
      };%
      \matrix [above of=facets] (vertices) {%
        \node [point,Aclos] (hal) {$\alpha$};
        &
        \node [point,Aclos] (hbe) {$\beta$};
        &
        \node [point,Aclos] (hga) {$\gamma$};
        &
        \node [point,deltastar] (hde) {$\delta$};
        \\
        };%
        \draw[arrow] (he) -- (hal);
        \draw[arrow] (hal) -- (he);
        \draw[arrow]             (ha) -- (hA);
        \draw[arrow,Acone,Aclos] (hA) -- (ha);
        \draw[arrow]             (hb) -- (hA);
        \draw[arrow,Acone,Aclos] (hA) -- (hb);
        \draw[arrow]             (hc) -- (hA);
        \draw[arrow,Acone,Aclos] (hA) -- (hc);
        \draw[arrow] (hB) -- (hc);
        \draw[arrow] (hc) -- (hB);
        \draw[arrow]           (hB) -- (hd);
        \draw[arrow,deltastar] (hd) -- (hB);
        \draw[arrow]           (hB) -- (he);
        \draw[arrow,deltastar] (he) -- (hB);
        \draw[arrow]       (hal) -- (ha);
        \draw[arrow,Aclos] (ha) -- (hal);
        \draw[arrow]       (hbe) -- (ha);
        \draw[arrow,Aclos] (ha) -- (hbe);
        \draw[arrow]       (hbe) -- (hb);
        \draw[arrow,Aclos] (hb) -- (hbe);
        \draw[arrow]       (hga) -- (hb);
        \draw[arrow]       (hga) -- (hc);
        \draw[arrow,Aclos] (hb) -- (hga);
        \draw[arrow]       (hal) -- (hc);
        \draw[arrow,Aclos] (hc) -- (hga);
        \draw[arrow] (hd) -- (hga);
        \draw[arrow,Aclos] (hc) -- (hal);
        \draw[arrow] (hga) -- (hd);
        \draw[arrow]                     (hd) -- (hde);
        \draw[arrow,deltasupp,deltastar] (hde) -- (hd);
        \draw[arrow]                     (he) -- (hde);
        \draw[arrow,deltasupp,deltastar] (hde) -- (he);
      \end{tikzpicture}

          };%
        \end{scope}
      \end{tikzpicture}
      \caption{%
        $\clos(A)$ / {\bf \protect\pd{DM}{DMPlexGetTransitiveClosure}(useCone=PETSC\_TRUE)}.%
      }%
      \label{fig:twotriangle:clos}%
  \end{figure}
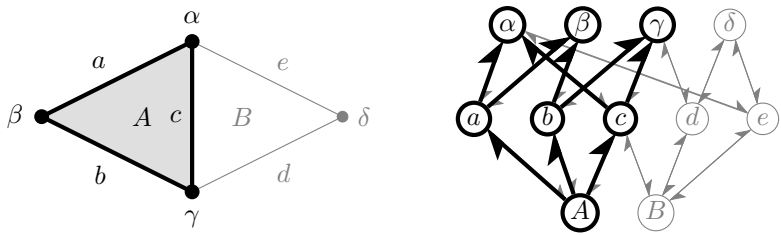

  \begin{figure}\centering
    \begin{tikzpicture}
      \begin{scope}[local bounding box=box]
        \tikzset{deltasupp/.append style={black,ultra thick}}
        \node (deltasupp) {%
          \begin{tikzpicture}
  \usetikzlibrary{thplex}
  \tikzset{Acone/.prefix style={}}
  \tikzset{Aclos/.prefix style={}}
  \tikzset{Aclosfill/.prefix style={}}
  \tikzset{deltasupp/.prefix style={}}
  \tikzset{deltastar/.prefix style={}}
  \tikzset{deltastarfill/.prefix style={}}
  \coordinate (al) at ( 0, 1);
  \coordinate (be) at (-2, 0);
  \coordinate (ga) at ( 0,-1);
  \coordinate (de) at ( 2, 0);
  \fill [cell,Aclos,Aclosfill] (al) -- (be) -- (ga) -- cycle
  node at (barycentric cs:al=1,be=1,ga=1) {$A$};
  \fill [cell,deltastar,deltastarfill] (al) -- (ga) -- (de) -- cycle
  node at (barycentric cs:al=1,ga=1,de=1) {$B$};
  \draw [facet,Acone,Aclos] (al) -- node [above left] {$a$} (be);
  \draw [facet,Acone,Aclos] (be) -- node [below left] {$b$} (ga);
  \draw [facet,Acone,Aclos] (ga) -- node [left] {$c$} (al);
  \draw [facet,deltasupp,deltastar] (ga) -- node [below right] {$d$} (de);
  \draw [facet,deltasupp,deltastar] (de) -- node [above right] {$e$} (al);
  \node at (al) [vertex,Aclos,label={[vertexlabel,Aclos]above:$\alpha$}] {};
  \node at (be) [vertex,Aclos,label={[vertexlabel,Aclos]left:$\beta$}] {};
  \node at (ga) [vertex,Aclos,label={[vertexlabel,Aclos]below:$\gamma$}] {};
  \node at (de) [vertex,deltastar,label={[vertexlabel,deltastar]right:$\delta$}] {};
\end{tikzpicture}
        };%
        \node [right=2cm,right of=deltasupp,anchor=west] (deltasupphasse) {%
          \begin{tikzpicture}
  \usetikzlibrary{thplex}
  \tikzset{Acone/.prefix style={}}
  \tikzset{Aclos/.prefix style={}}
  \tikzset{deltasupp/.prefix style={}}
  \tikzset{deltastar/.prefix style={}}
  \matrix (cells) {%
    \node [point,Aclos] (hA) {$A$};
    &
    \node [point,deltastar] (hB) {$B$};
    \\
    };%
    \matrix [above of=cells] (facets) {%
      \node [point,Acone,Aclos] (ha) {$a$};
      &
      \node [point,Acone,Aclos] (hb) {$b$};
      &
      \node [point,Acone,Aclos] (hc) {$c$};
      &
      \node [point,deltasupp,deltastar] (hd) {$d$};
      &
      \node [point,deltasupp,deltastar] (he) {$e$};
      \\
      };%
      \matrix [above of=facets] (vertices) {%
        \node [point,Aclos] (hal) {$\alpha$};
        &
        \node [point,Aclos] (hbe) {$\beta$};
        &
        \node [point,Aclos] (hga) {$\gamma$};
        &
        \node [point,deltastar] (hde) {$\delta$};
        \\
        };%
        \draw[arrow] (he) -- (hal);
        \draw[arrow] (hal) -- (he);
        \draw[arrow]             (ha) -- (hA);
        \draw[arrow,Acone,Aclos] (hA) -- (ha);
        \draw[arrow]             (hb) -- (hA);
        \draw[arrow,Acone,Aclos] (hA) -- (hb);
        \draw[arrow]             (hc) -- (hA);
        \draw[arrow,Acone,Aclos] (hA) -- (hc);
        \draw[arrow] (hB) -- (hc);
        \draw[arrow] (hc) -- (hB);
        \draw[arrow]           (hB) -- (hd);
        \draw[arrow,deltastar] (hd) -- (hB);
        \draw[arrow]           (hB) -- (he);
        \draw[arrow,deltastar] (he) -- (hB);
        \draw[arrow]       (hal) -- (ha);
        \draw[arrow,Aclos] (ha) -- (hal);
        \draw[arrow]       (hbe) -- (ha);
        \draw[arrow,Aclos] (ha) -- (hbe);
        \draw[arrow]       (hbe) -- (hb);
        \draw[arrow,Aclos] (hb) -- (hbe);
        \draw[arrow]       (hga) -- (hb);
        \draw[arrow]       (hga) -- (hc);
        \draw[arrow,Aclos] (hb) -- (hga);
        \draw[arrow]       (hal) -- (hc);
        \draw[arrow,Aclos] (hc) -- (hga);
        \draw[arrow] (hd) -- (hga);
        \draw[arrow,Aclos] (hc) -- (hal);
        \draw[arrow] (hga) -- (hd);
        \draw[arrow]                     (hd) -- (hde);
        \draw[arrow,deltasupp,deltastar] (hde) -- (hd);
        \draw[arrow]                     (he) -- (hde);
        \draw[arrow,deltasupp,deltastar] (hde) -- (he);
      \end{tikzpicture}

        };%
      \end{scope}
    \end{tikzpicture}
    \caption{$\supp(\delta)$ / {\bf \protect\pd{DM}{DMPlexGetSupport}()}.}%
    \label{fig:twotriangle:supp}%
  \end{figure}
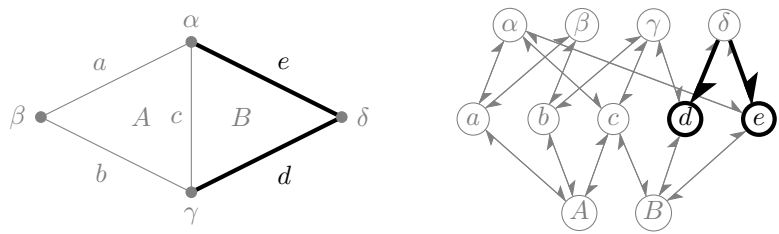

  \begin{figure}\centering
    \begin{tikzpicture}
      \begin{scope}[local bounding box=box]
        \tikzset{deltastar/.append style={black,ultra thick}}
        \tikzset{deltastarfill/.append style={fill=gray!25!white}}
        \node (deltastar) {%
          \begin{tikzpicture}
  \usetikzlibrary{thplex}
  \tikzset{Acone/.prefix style={}}
  \tikzset{Aclos/.prefix style={}}
  \tikzset{Aclosfill/.prefix style={}}
  \tikzset{deltasupp/.prefix style={}}
  \tikzset{deltastar/.prefix style={}}
  \tikzset{deltastarfill/.prefix style={}}
  \coordinate (al) at ( 0, 1);
  \coordinate (be) at (-2, 0);
  \coordinate (ga) at ( 0,-1);
  \coordinate (de) at ( 2, 0);
  \fill [cell,Aclos,Aclosfill] (al) -- (be) -- (ga) -- cycle
  node at (barycentric cs:al=1,be=1,ga=1) {$A$};
  \fill [cell,deltastar,deltastarfill] (al) -- (ga) -- (de) -- cycle
  node at (barycentric cs:al=1,ga=1,de=1) {$B$};
  \draw [facet,Acone,Aclos] (al) -- node [above left] {$a$} (be);
  \draw [facet,Acone,Aclos] (be) -- node [below left] {$b$} (ga);
  \draw [facet,Acone,Aclos] (ga) -- node [left] {$c$} (al);
  \draw [facet,deltasupp,deltastar] (ga) -- node [below right] {$d$} (de);
  \draw [facet,deltasupp,deltastar] (de) -- node [above right] {$e$} (al);
  \node at (al) [vertex,Aclos,label={[vertexlabel,Aclos]above:$\alpha$}] {};
  \node at (be) [vertex,Aclos,label={[vertexlabel,Aclos]left:$\beta$}] {};
  \node at (ga) [vertex,Aclos,label={[vertexlabel,Aclos]below:$\gamma$}] {};
  \node at (de) [vertex,deltastar,label={[vertexlabel,deltastar]right:$\delta$}] {};
\end{tikzpicture}
        };%
        \node [right=2cm,right of=deltastar,anchor=west] (deltastarhasse) {%
          \begin{tikzpicture}
  \usetikzlibrary{thplex}
  \tikzset{Acone/.prefix style={}}
  \tikzset{Aclos/.prefix style={}}
  \tikzset{deltasupp/.prefix style={}}
  \tikzset{deltastar/.prefix style={}}
  \matrix (cells) {%
    \node [point,Aclos] (hA) {$A$};
    &
    \node [point,deltastar] (hB) {$B$};
    \\
    };%
    \matrix [above of=cells] (facets) {%
      \node [point,Acone,Aclos] (ha) {$a$};
      &
      \node [point,Acone,Aclos] (hb) {$b$};
      &
      \node [point,Acone,Aclos] (hc) {$c$};
      &
      \node [point,deltasupp,deltastar] (hd) {$d$};
      &
      \node [point,deltasupp,deltastar] (he) {$e$};
      \\
      };%
      \matrix [above of=facets] (vertices) {%
        \node [point,Aclos] (hal) {$\alpha$};
        &
        \node [point,Aclos] (hbe) {$\beta$};
        &
        \node [point,Aclos] (hga) {$\gamma$};
        &
        \node [point,deltastar] (hde) {$\delta$};
        \\
        };%
        \draw[arrow] (he) -- (hal);
        \draw[arrow] (hal) -- (he);
        \draw[arrow]             (ha) -- (hA);
        \draw[arrow,Acone,Aclos] (hA) -- (ha);
        \draw[arrow]             (hb) -- (hA);
        \draw[arrow,Acone,Aclos] (hA) -- (hb);
        \draw[arrow]             (hc) -- (hA);
        \draw[arrow,Acone,Aclos] (hA) -- (hc);
        \draw[arrow] (hB) -- (hc);
        \draw[arrow] (hc) -- (hB);
        \draw[arrow]           (hB) -- (hd);
        \draw[arrow,deltastar] (hd) -- (hB);
        \draw[arrow]           (hB) -- (he);
        \draw[arrow,deltastar] (he) -- (hB);
        \draw[arrow]       (hal) -- (ha);
        \draw[arrow,Aclos] (ha) -- (hal);
        \draw[arrow]       (hbe) -- (ha);
        \draw[arrow,Aclos] (ha) -- (hbe);
        \draw[arrow]       (hbe) -- (hb);
        \draw[arrow,Aclos] (hb) -- (hbe);
        \draw[arrow]       (hga) -- (hb);
        \draw[arrow]       (hga) -- (hc);
        \draw[arrow,Aclos] (hb) -- (hga);
        \draw[arrow]       (hal) -- (hc);
        \draw[arrow,Aclos] (hc) -- (hga);
        \draw[arrow] (hd) -- (hga);
        \draw[arrow,Aclos] (hc) -- (hal);
        \draw[arrow] (hga) -- (hd);
        \draw[arrow]                     (hd) -- (hde);
        \draw[arrow,deltasupp,deltastar] (hde) -- (hd);
        \draw[arrow]                     (he) -- (hde);
        \draw[arrow,deltasupp,deltastar] (hde) -- (he);
      \end{tikzpicture}

        };%
      \end{scope}
    \end{tikzpicture}
    \caption{%
      $\Star(\delta)$ / {\bf
      \protect\pd{DM}{DMPlexGetTransitiveClosure}(useCone=PETSC\_FALSE)}.%
    }%
    \label{fig:twotriangle:star}%
  \end{figure}
}%

\subsection{The reference element and element maps}

We assume a Ciarlet reference finite element $(\hat{K}, P(\hat{K}),
\hat{\Sigma})$ (reference cell, space, and dual basis) is specified and a
domain $\Omega$ is triangulated into a mesh of $N_K$ cells, with cell $K_i$
being the image of $\hat{K}$ under a smooth embedding $\varphi_i:
\overline{\hat{K}} \to \overline{\Omega}$.  A finite-dimensional subspace
$V_h$ of a function space $V(\Omega)$ is then specified as the set of
functions $v\in V$ such that the pullback $\varphi_i^* v:=v\circ \varphi_i$ is
in $P(\hat{K})$ for each $\varphi_i$.  (Discretizations of
$H^{\Curl}(\Omega)$-{} and $H^{\Div}(\Omega)$-conforming spaces are often
pulled back onto reference elements using covariant and contravariant Piola
transformations, which have minor implications discussed \cref{sec:curldiv}.)
The adjoint of the pullback is the pushforward
$(\varphi_{i*}\sigma)(v):=\sigma(\varphi_i^*v)$: it pushes $\hat{\Sigma}$
forward onto a set of functionals in $V_h^*$,
\begin{equation}\label{eq:pushforward}
  \Sigma_i := \varphi_{i*}\hat{\Sigma}.
\end{equation}

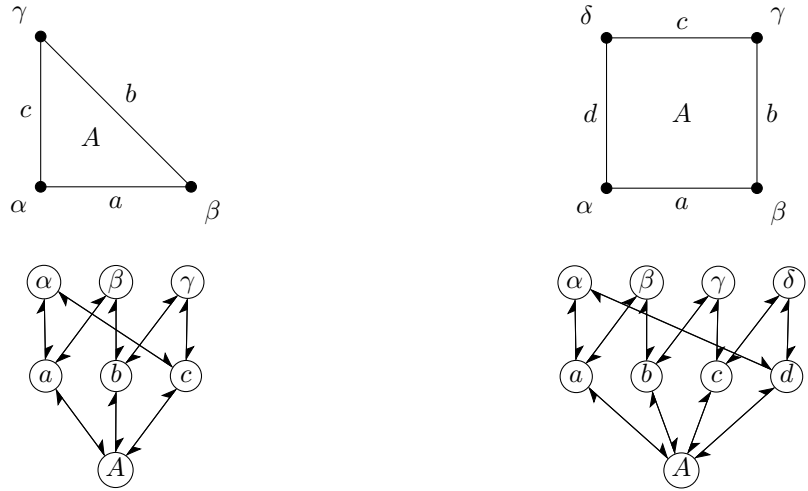
\begin{figure}\centering%
  \begin{minipage}{0.49\textwidth}\centering%
    \begin{tikzpicture}
      \begin{scope}[local bounding box=box]
        \node (triangle) {%
          \begin{tikzpicture}
  \usetikzlibrary{thplex}
  \coordinate (al) at (-1,-1); \node at (al) [vertex,label={[vertexlabel]below left:$\alpha$}] {};
  \coordinate (be) at ( 1,-1); \node at (be) [vertex,label={[vertexlabel]below right:$\beta$}] {};
  \coordinate (ga) at (-1, 1); \node at (ga) [vertex,label={[vertexlabel]above left:$\gamma$}] {};
  \fill [cell] (al) -- (be) -- (ga) -- cycle node at (barycentric cs:al=1,be=1,ga=1) {$A$};
  \draw [facet] (al) -- node [below]       {$a$} (be);
  \draw [facet] (be) -- node [above right] {$b$} (ga);
  \draw [facet] (ga) -- node [left]        {$c$} (al);
\end{tikzpicture}

        };%
        \node [below=0.5cm,below of=triangle, anchor=north] (trianglehasse) {%
          \begin{tikzpicture}
  \matrix (cells) {%
    \node [point] (hA) {$A$}; \\
  };%
  \matrix (facets) [above of=cells]{%
    \node [point] (ha) {$a$};
    &
    \node [point] (hb) {$b$};
    &
    \node [point] (hc) {$c$};
    \\
  };%
  \matrix (vertices) [above of=facets]{%
    \node [point] (hal) {$\alpha$};
    &
    \node [point] (hbe) {$\beta$};
    &
    \node [point] (hga) {$\gamma$};
    \\
  };%
  \draw[arrow] (hA) -- (ha);
  \draw[arrow] (ha) -- (hA);
  \draw[arrow] (hA) -- (hb);
  \draw[arrow] (hb) -- (hA);
  \draw[arrow] (hA) -- (hc);
  \draw[arrow] (hc) -- (hA);
  \draw[arrow] (ha) -- (hal);
  \draw[arrow] (hal) -- (ha);
  \draw[arrow] (ha) -- (hbe);
  \draw[arrow] (hbe) -- (ha);
  \draw[arrow] (hb) -- (hbe);
  \draw[arrow] (hbe) -- (hb);
  \draw[arrow] (hb) -- (hga);
  \draw[arrow] (hga) -- (hb);
  \draw[arrow] (hc) -- (hga);
  \draw[arrow] (hga) -- (hc);
  \draw[arrow] (hc) -- (hal);
  \draw[arrow] (hal) -- (hc);
\end{tikzpicture}
        };%
      \end{scope}
    \end{tikzpicture}%
  \end{minipage}
  \begin{minipage}{0.49\textwidth}\centering%
    \begin{tikzpicture}
      \begin{scope}[local bounding box=box]
        \node (quad) {%
          \begin{tikzpicture}
  \usetikzlibrary{thplex}
  \coordinate (al) at (-1,-1); \node at (al) [vertex,label={[vertexlabel]below left:$\alpha$}]  {};
  \coordinate (be) at ( 1,-1); \node at (be) [vertex,label={[vertexlabel]below right:$\beta$}]  {};
  \coordinate (ga) at ( 1, 1); \node at (ga) [vertex,label={[vertexlabel]above right:$\gamma$}] {};
  \coordinate (de) at (-1, 1); \node at (de) [vertex,label={[vertexlabel]above left:$\delta$}]  {};
  \fill [cell] (al) -- (be) -- (ga) -- (de) -- cycle;
  \node at (barycentric cs:al=1,be=1,ga=1,de=1) {$A$};
  \draw [facet] (al) -- node [below] {$a$} (be);
  \draw [facet] (be) -- node [right] {$b$} (ga);
  \draw [facet] (ga) -- node [above] {$c$} (de);
  \draw [facet] (de) -- node [left]  {$d$} (al);
\end{tikzpicture}
        };%
        \node [below=0.5cm,below of=quad,anchor=north] (quadhasse) {%
          \begin{tikzpicture}
  \usetikzlibrary{thplex}
  \matrix (cells) {%
    \node [point] (hA) {$A$}; \\
  };%
  \matrix (facets) [above of=cells]{%
    \node [point] (ha) {$a$};
    &
    \node [point] (hb) {$b$};
    &
    \node [point] (hc) {$c$};
    &
    \node [point] (hd) {$d$};
    \\
  };%
  \matrix (vertices) [above of=facets]{%
    \node [point] (hal) {$\alpha$};
    &
    \node [point] (hbe) {$\beta$};
    &
    \node [point] (hga) {$\gamma$};
    &
    \node [point] (hde) {$\delta$};
    \\
  };%
  \draw[arrow] (hA) -- (ha);
  \draw[arrow] (ha) -- (hA);
  \draw[arrow] (hA) -- (hb);
  \draw[arrow] (hb) -- (hA);
  \draw[arrow] (hA) -- (hc);
  \draw[arrow] (hc) -- (hA);
  \draw[arrow] (hA) -- (hd);
  \draw[arrow] (hd) -- (hA);
  \draw[arrow] (ha) -- (hal);
  \draw[arrow] (hal) -- (ha);
  \draw[arrow] (ha) -- (hbe);
  \draw[arrow] (hbe) -- (ha);
  \draw[arrow] (hb) -- (hbe);
  \draw[arrow] (hbe) -- (hb);
  \draw[arrow] (hb) -- (hga);
  \draw[arrow] (hga) -- (hb);
  \draw[arrow] (hc) -- (hga);
  \draw[arrow] (hga) -- (hc);
  \draw[arrow] (hc) -- (hde);
  \draw[arrow] (hde) -- (hc);
  \draw[arrow] (hd) -- (hde);
  \draw[arrow] (hde) -- (hd);
  \draw[arrow] (hd) -- (hal);
  \draw[arrow] (hal) -- (hd);
\end{tikzpicture}
        };%
      \end{scope}
    \end{tikzpicture}%
  \end{minipage}
  \caption{%
    The reference triangle, reference quadrilateral, and the Hasse diagrams of
    their CW-complexes.
  }%
  \label{fig:refcomplex}%
\end{figure}

To the conventional triplet $(\hat{K}, P(\hat{K}), \hat{\Sigma})$, we add a
reference CW-complex $\hat{S}$, which decomposes the closure of $\hat{K}$
(\cref{fig:refcomplex}).

\subsection{The finite element method for conformal meshes in DMPlex}

We now list three assumptions that are implicit in most finite element
discretizations for conformal meshes, and thus in the data structures and
functions that DMPlex uses to implement the finite element method. 
Making these typically implicit assumptions explicit will help to explain the
extension of DMPlex to non-conformal meshes in \cref{sec:nonconformal}.

First, we assume that the reference complex $\hat{S}$ also decomposes the dual
basis $\hat{\Sigma}$, in that the shape function associated with each
$\sigma\in\hat{\Sigma}$ is supported in the star of a point in $\hat{S}$.  We
formalize this as assumption I.
\begin{itemize}
  \item[I.] 
    For each $\sigma_k\in\hat{\Sigma}$ there is a point $p\in\hat{S}$ such
    that, if $\psi_k\in P(\hat{K})$ is $\sigma_k$'s shape function
    ($\sigma_j(\psi_k)=\delta_{jk}$), then $\supp(\psi_k) = \cup
    \Star(p)$.
\end{itemize}
Assumption I is satisfied by essentially all finite elements: it allows for
the definition of compactly supported basis functions of $V_h$.  We will refer
to the reference functionals associated with $p\in \hat{S}$ as
$\hat{\Sigma}^p$, so that $\hat{\Sigma} = \cup_{p\in\hat{S}} \hat{\Sigma}^p$,
and to the pushforward of those functionals under $\varphi_i$ as
$\Sigma_i^p:=\varphi_{i*}\hat{\Sigma}^p$, so that $\Sigma_i =
\cup_{p\in\hat{S}} \Sigma_i^p$.

Second, we assume that the embeddings of neighboring cells are
compatible, in that the traces of their approximation spaces ``line up'' so
that $H^1(\Omega)$ functions can be constructed, which we formalize
as assumption II.
\begin{enumerate}
  \item[II.]
    If $C:=\overline{K_i}\cap \overline{K_j}\neq \emptyset$, then $\psi\in
    P(\varphi_j^{-1}(C))\Rightarrow \varphi_i^*\varphi_j^{-*} \psi \in
    P(\varphi_i^{-1}(C))$ (where $\varphi_j^{-*}:= (\varphi_j^*)^{-1}$ and
    $P(X)$ is the trace space of $P(\hat{K})$ on $X \subset
    \overline{\hat{K}}$).
\end{enumerate}

Finally, we assume that the dual bases of adjacent cells are compatible, in
that the mappings of adjacent cells push functionals forward on top of each
other, which we formalize as assumption III.
\begin{enumerate}
  \item[III.]
    If $p,q\in \hat{S}$ and there are adjacent cells $K_i$ and $K_j$ such that
    $\varphi_i(p) = \varphi_j(q)$, then there is a permutation $M$ such that
    $\Sigma_i^p = M \Sigma_j^q$.
\end{enumerate}
The permutations typically encode the symmetries of the polytopes in
$\hat{S}$, e.g., reversal for edges, and dihedral symmetries for faces.

For each vector $v\in V_h$ and each element $K_i$, we need to be able to
evaluate $\Sigma_i(v)$.  Given a choice of basis $W$ for $V_h^*$, each element
has a \emph{restriction matrix} $R_i$ such that $\Sigma_i(v) = R_i W (v)$.
For a conformal mesh, assumptions I, II, and III allow for a \emph{global
nodal basis} to be defined for $V_h^*$: a basis $W$ that is the union of the
pushforward dual bases,
\begin{equation}
  W := \cup_{i=1}^{N_K} \cup_{p\in\hat{S}} \Sigma_i^p.
\end{equation}
We may also think of $W$ as being decomposed into the functional associated with
points in $S$,
\begin{equation}
  W = \cup_{s\in S} \Sigma^s,
\end{equation}
where $\Sigma^s := \Sigma_i^p$ (up to a permutation) if $\varphi_i(p)=s$.  With
a global nodal basis, the restriction matrix $R_i$ for each cell $K_i$ is a
binary matrix, and there is a subset $W_i$ of $W$ such that $\Sigma_i = R_i W
= W_i$.  

Here we see the utility of representing a conformal mesh as a CW-complex $S$.
Given a map $G:S\to 2^W$ from each point in $S$ to its set of associated
functionals in $W$, we can compute $W_i$ as $G(\clos(K_i))$, since
$\clos(K_i)$ is the image of the reference complex $\hat{S}$ under
$\varphi_i$. 

In DMPlex, the map $G$ is represented by a {\bf\pd{IS}{PetscSection}}.  For a
typical finite element, the number of functionals associated with a point $p\in
S$ is a function only of $p$'s topological dimension, so that $G$ can be
calculated purely from the sizes of the strata of $S$: if a DMPlex has been
given a finite element object (PetscFE), it constructs $G$ automatically, and
makes it available by {\bf\pd{DM}{DMGetDefaultGlobalSection}()}.  The set
$\clos(K_i)$ can be constructed with
{\bf\pd{DM}{DMPlexGetTransitiveClosure}()}, which is used to construct
$\Sigma_i(v)$ on a vector $v\in V_h^*$ in the function
{\bf\pd{DM}{DMPlexVecGetClosure}()}.  This function is called within tight,
performance critical loops when computing residuals or calculating Jacobians
(which we illustrate in a prototypical residual evaluation function in
\cref{alg:feresidual}), so when considering representations of non-conformal
meshes in DMPlex, we chose to avoid those that would require modifications at
this level of granularity.

\renewcommand{\algorithmcfname}{EXAMPLE}
\begin{algorithm}
  \hypersetup{urlbordercolor={1 1 1}}
  \caption{finite element residual ${\tt r} = f({\tt v})$ using DMPlex ({\tt
  dm})}%
  \label{alg:feresidual}%

  \pd{Sys}{PetscInt} c, cSize, cStart, cEnd;

  \pd{Vec}{Vec} \ \ \ \ \ vLocal, rLocal;

  \null

  \tcc{Get vectors for the local representations}

  \ierr \fn{\pd{DM}{DMGetLocalVector}}(dm,\&vLocal);\chkerr

  \ierr \fn{\pd{DM}{DMGetLocalVector}}(dm,\&rLocal);\chkerr

  \tcc{Get the support of v on the local subdomain of this MPI process}

  \ierr \fn{\pd{DM}{DMGlobalToLocalBegin}}(dm,v,INSERT\_VALUES,vLocal);\chkerr

  \ierr \fn{\pd{DM}{DMGlobalToLocalEnd}} \ (dm,v,INSERT\_VALUES,vLocal);\chkerr%
  \label{line:globaltolocal}

  \tcc{Get the range of cell indices (cells are the lowest stratum)}

  \ierr \fn{\pd{DM}{DMPlexGetHeightStratum}}(dm,0,\&cStart,\&cEnd);\chkerr

  \If{(cEnd > cStart)}{%

    PetscScalar *vElem, *rElem;

    \tcc{Get the size of the element dual space $\hat{\Sigma}$}

    \ierr \fn{\pd{DM}{DMPlexGetVecClosure}}(dm,\nll,vLocal,cStart,\&cSize,\nll);\chkerr

    \tcc{Get workspace arrays}

    \ierr \fn{\pd{DM}{DMPlexGetWorkArray}}(dm,cSize,PETSC\_SCALAR,\&vElem);\chkerr

    \ierr \fn{DMPlexGetWorkArray}(dm,cSize,PETSC\_SCALAR,\&rElem);\chkerr

    \tcc{Compute the local residual}

    \ierr \fn{\pd{Vec}{VecSet}}(rLocal,0.0);\chkerr

    \For{(c = cStart; c < cEnd; c++)}{%
      
      \tcc{Get the restriction of v to cell c}

      \ierr \fn{\pd{DM}{DMPlexVecGetClosure}}(dm,\nll,vLocal,c,\&cSize,\&vElem);\chkerr

      \tcc{Compute the element contribution to the residual}

      \tcc{\dots\ [rElem = f(vElem)]}

      \tcc{Sum the element residual into the local residual vector}

      \ierr \fn{\pd{DM}{DMPlexVecSetClosure}}(dm,\nll,rLocal,c,rElem,ADD\_VALUES);\chkerr

    }%

    \tcc{Free workspace}

    \ierr \fn{\pd{DM}{DMPlexRestoreWorkArray}}(dm,cSize,PETSC\_SCALAR,\&vElem);\chkerr

    \ierr \fn{\pd{DM}{DMPlexRestoreWorkArray}}(dm,cSize,PETSC\_SCALAR,\&rElem);\chkerr

  }%

  \tcc{Sum process contributions into r}

  \ierr \fn{\pd{Vec}{VecSet}}(r,0.0);\chkerr

  \ierr \fn{\pd{DM}{DMLocalToGlobalBegin}}(dm,rLocal,ADD\_VALUES,r);\chkerr

  \ierr \fn{\pd{DM}{DMLocalToGlobalEnd}} \ (dm,rLocal,ADD\_VALUES,r);\chkerr
  \label{line:localtoglobal}

  \tcc{Free local vectors}

  \ierr \fn{\pd{DM}{DMRestoreLocalVector}}(dm,\&vLocal);\chkerr

  \ierr \fn{\pd{DM}{DMRestoreLocalVector}}(dm,\&rLocal);\chkerr

\end{algorithm}

\section{Non-conformal meshes}
\label{sec:nonconformal}

In this section we describe the way non-conformal meshes can now be
represented in DMPlex, and we describe a general approach to computing with
finite elements on these meshes.

\subsection{Representing non-conformal meshes in DMPlex}

The representation of non-conformal meshes that has been added to DMPlex is
limited to hierarchical non-conformal meshes.  By hierarchical we mean that
two mesh points $p,q\in S$ overlap only if one is a superset of the other,
\begin{equation}
  (p\cap q \neq \emptyset) \Rightarrow ((p\subseteq q) \vee (q \subseteq p)).
\end{equation}
Constructing function spaces on non-conformal meshes that are not hierarchical
is more complicated, and not considered here.

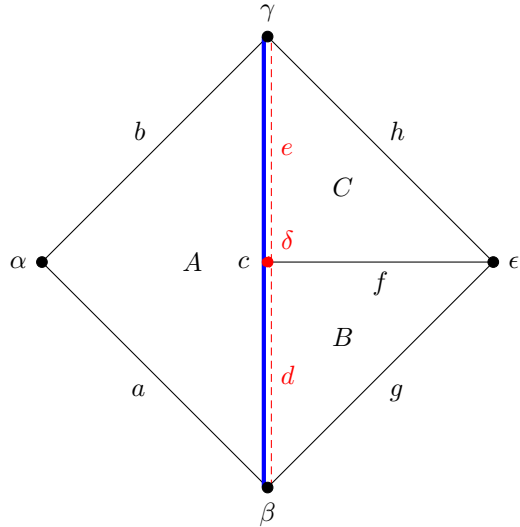
\begin{figure}\centering
  \tikzset{parent/.append style={blue,ultra thick}}
  \tikzset{child/.append style={red,densely dashed}}
  \tikzset{childlabel/.append style={red}}
  \begin{tikzpicture}
  \usetikzlibrary{thplex}
  \tikzset{parent/.prefix style={}}
  \tikzset{child/.prefix style={}}
  \tikzset{childlabel/.prefix style={}}
  \tikzset{parentlabel/.prefix style={}}
  \tikzset{star/.prefix style={}}
  \tikzset{starfill/.prefix style={}}

  \coordinate (al) at (-3, 0);
  \coordinate (be) at ( 0,-3);
  \coordinate (ga) at ( 0, 3);
  \coordinate (ep) at ( 3, 0);
  \coordinate (de) at ( 0, 0);
  \fill [cell,starfill] (al) -- (be) -- (ga) -- cycle
    node at (barycentric cs:al=1,be=1,ga=1) {$A$};
  \fill [cell] (be) -- (ep) -- (de) -- cycle
    node at (barycentric cs:be=1,ep=1,de=1) {$B$};
  \fill [cell,starfill] (ep) -- (ga) -- (de) -- cycle
    node at (barycentric cs:ep=1,ga=1,de=1) {$C$};
  \draw [facet] (al) -- node [below left] {$a$} (be);
  \draw [facet,star] (ga) -- node [above left] {$b$} (al);
  \draw [facet,shift left=0.5mm,parent] (be) -- node [facetlabel,parentlabel,left,xshift=-1mm] {$c$} (ga);
  \draw [facet,shift right=0.5mm,child] (be) -- node [facetlabel,childlabel,right,xshift=0.5mm] {$d$} (de);
  \draw [facet,shift right=0.5mm,star,child] (de) -- node [facetlabel,childlabel,right,xshift=0.5mm] {$e$} (ga);
  \draw [facet] (ep) -- node [below] {$f$} (de);
  \draw [facet] (be) -- node [below right] {$g$} (ep);
  \draw [facet,star] (ep) -- node [above right] {$h$} (ga);
  \node at (al) [vertex,label={[vertexlabel]left:$\alpha$}] {};
  \node at (be) [vertex,label={[vertexlabel]below:$\beta$}] {};
  \node at (ga) [vertex,star,label={[vertexlabel,star]above:$\gamma$}] {};
  \node at (de) [vertex,child,label={[vertexlabel,childlabel]above right:$\delta$}] {};
  \node at (ep) [vertex,label={[vertexlabel]right:$\epsilon$}] {};

\end{tikzpicture}

  \caption{A simple non-conformal mesh.}
  \label{fig:mesh}
\end{figure}

In \cref{fig:mesh} we show a simple three-triangle mesh with a non-conformal
interface between triangle $A$ on one side and triangles $B$ and $C$ on the
other, and we have also labeled all of the edges and vertices of the
triangles.  The question when representing this mesh in DMPlex is how to treat
the edge $c$ and the edges $d$, $e$, and vertex $\delta$ that overlap it.

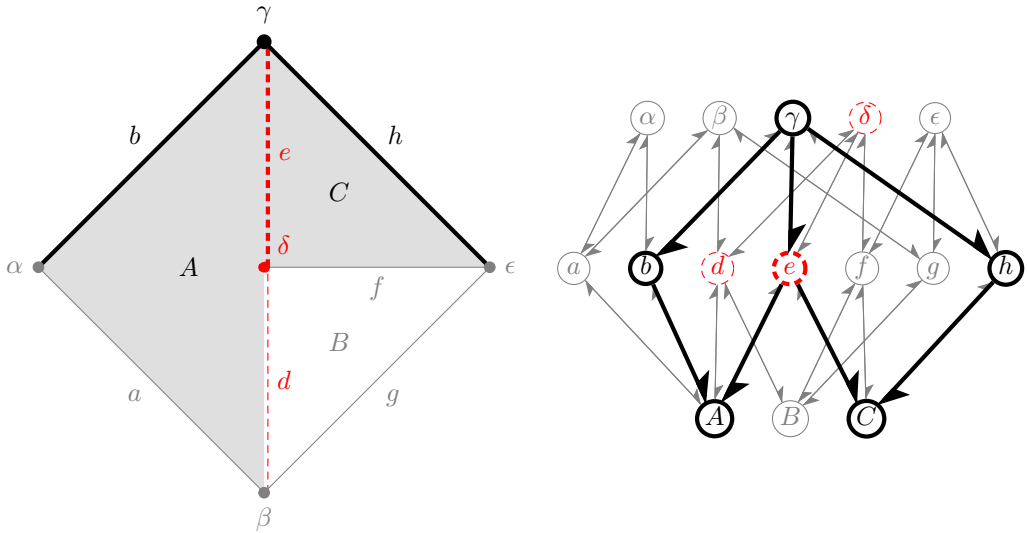
\begin{figure}\centering
  \tikzset{point/.append style={gray}}
  \tikzset{vertex/.append style={gray}}
  \tikzset{vertexlabel/.append style={gray}}
  \tikzset{facet/.append style={gray}}
  \tikzset{cell/.append style={gray}}
  \tikzset{every picture/.append style={draw=gray,fill=gray}}
  \tikzset{parent/.append style={draw=none,fill=none}}
  \tikzset{parentlabel/.append style={opacity=0.}}
  \tikzset{child/.append style={red,densely dashed}}
  \tikzset{childlabel/.append style={red}}
  \tikzset{star/.append style={black,ultra thick}}
  \tikzset{starfill/.append style={fill=gray!25!white,text=black}}
  \begin{minipage}{0.49\textwidth}\centering
    \begin{tikzpicture}
  \usetikzlibrary{thplex}
  \tikzset{parent/.prefix style={}}
  \tikzset{child/.prefix style={}}
  \tikzset{childlabel/.prefix style={}}
  \tikzset{parentlabel/.prefix style={}}
  \tikzset{star/.prefix style={}}
  \tikzset{starfill/.prefix style={}}

  \coordinate (al) at (-3, 0);
  \coordinate (be) at ( 0,-3);
  \coordinate (ga) at ( 0, 3);
  \coordinate (ep) at ( 3, 0);
  \coordinate (de) at ( 0, 0);
  \fill [cell,starfill] (al) -- (be) -- (ga) -- cycle
    node at (barycentric cs:al=1,be=1,ga=1) {$A$};
  \fill [cell] (be) -- (ep) -- (de) -- cycle
    node at (barycentric cs:be=1,ep=1,de=1) {$B$};
  \fill [cell,starfill] (ep) -- (ga) -- (de) -- cycle
    node at (barycentric cs:ep=1,ga=1,de=1) {$C$};
  \draw [facet] (al) -- node [below left] {$a$} (be);
  \draw [facet,star] (ga) -- node [above left] {$b$} (al);
  \draw [facet,shift left=0.5mm,parent] (be) -- node [facetlabel,parentlabel,left,xshift=-1mm] {$c$} (ga);
  \draw [facet,shift right=0.5mm,child] (be) -- node [facetlabel,childlabel,right,xshift=0.5mm] {$d$} (de);
  \draw [facet,shift right=0.5mm,star,child] (de) -- node [facetlabel,childlabel,right,xshift=0.5mm] {$e$} (ga);
  \draw [facet] (ep) -- node [below] {$f$} (de);
  \draw [facet] (be) -- node [below right] {$g$} (ep);
  \draw [facet,star] (ep) -- node [above right] {$h$} (ga);
  \node at (al) [vertex,label={[vertexlabel]left:$\alpha$}] {};
  \node at (be) [vertex,label={[vertexlabel]below:$\beta$}] {};
  \node at (ga) [vertex,star,label={[vertexlabel,star]above:$\gamma$}] {};
  \node at (de) [vertex,child,label={[vertexlabel,childlabel]above right:$\delta$}] {};
  \node at (ep) [vertex,label={[vertexlabel]right:$\epsilon$}] {};

\end{tikzpicture}

  \end{minipage}
  \begin{minipage}{0.49\textwidth}
    \begin{tikzpicture}[node distance=2cm]
  \usetikzlibrary{thplex}
  \tikzset{parent/.prefix style={}}
  \tikzset{child/.prefix style={}}
  \tikzset{star/.prefix style={}}

  \matrix (cells) {
    \node [point,star] (A) {$A$}; &
    \node [point] (B) {$B$}; &
    \node [point,star] (C) {$C$}; \\
  };
  \matrix [above of=cells] (facets) {
    \node [point] (a) {$a$}; &
    \node [point,star] (b) {$b$}; &
    \node [point,child] (d) {$d$}; &
    \node [point,star,child] (e) {$e$}; &
    \node [point] (f) {$f$}; &
    \node [point] (g) {$g$}; &
    \node [point,star] (h) {$h$}; \\
  };
  \matrix [above of=facets] (vertices) {
    \node [point] (al) {$\alpha$}; &
    \node [point] (be) {$\beta$};  &
    \node [point,star] (ga) {$\gamma$}; &
    \node [point,child] (de) {$\delta$}; &
    \node [point] (ep) {$\epsilon$}; \\
  };
  \draw[arrow] (A) -- (a);
  \draw[arrow] (a) -- (A);
  \draw[arrow] (A) -- (b);
  \draw[arrow] (A) -- (d);
  \draw[arrow] (d) -- (A);
  \draw[arrow] (A) -- (e);

  \draw[arrow] (B) -- (d);
  \draw[arrow] (d) -- (B);
  \draw[arrow] (B) -- (f);
  \draw[arrow] (f) -- (B);
  \draw[arrow] (B) -- (g);
  \draw[arrow] (g) -- (B);

  \draw[arrow] (C) -- (e);
  \draw[arrow] (C) -- (f);
  \draw[arrow] (f) -- (C);
  \draw[arrow] (C) -- (h);

  \draw[arrow] (a) -- (al);
  \draw[arrow] (al) -- (a);
  \draw[arrow] (a) -- (be);
  \draw[arrow] (be) -- (a);

  \draw[arrow] (b) -- (al);
  \draw[arrow] (al) -- (b);
  \draw[arrow] (b) -- (ga);

  \draw[arrow] (d) -- (be);
  \draw[arrow] (be) -- (d);
  \draw[arrow] (d) -- (de);
  \draw[arrow] (de) -- (d);

  \draw[arrow] (e) -- (de);
  \draw[arrow] (de) -- (e);
  \draw[arrow] (e) -- (ga);

  \draw[arrow] (f) -- (de);
  \draw[arrow] (de) -- (f);
  \draw[arrow] (f) -- (ep);
  \draw[arrow] (ep) -- (f);

  \draw[arrow] (g) -- (be);
  \draw[arrow] (be) -- (g);
  \draw[arrow] (g) -- (ep);
  \draw[arrow] (ep) -- (g);

  \draw[arrow] (h) -- (ga);
  \draw[arrow] (h) -- (ep);
  \draw[arrow] (ep) -- (h);
  \draw[arrow,star] (ga) -- (h);
  \draw[arrow,star] (ga) -- (e);
  \draw[arrow,star] (ga) -- (b);
  \draw[arrow,star] (h) -- (C);
  \draw[arrow,star] (b) -- (A);
  \draw[arrow,star] (e) -- (C);
  \draw[arrow,star] (e) -- (A);
\end{tikzpicture}

  \end{minipage}
  \caption{%
    A true CW-complex representation of \cref{fig:mesh}, demonstrating that
    $\Star(\gamma)$ does not include $B$.
  }%
  \label{fig:hassebad}
\end{figure}

The only way to represent this mesh as a true CW-complex (which cannot have
overlapping points) is to remove the long edge $c$ (\cref{fig:hassebad}), but
then the cell $A$ would not be considered a triangle: it would be considered a
degenerate quadrilateral, with $d$ and $e$ in the cone of $A$.  This is a poor
format for finite element computations for two reasons.  The first is that the
support of a basis function is no longer correlated with the star operator.  A
hat function centered at vertex $\gamma$, for instance, is non-zero on cell
$B$, but $B\not\in\Star(\gamma)$.  The second reason is that the shape of
$\clos(A)$ is not the same as for other triangles, so DMPlexVecGetClosure()
would require special handling to restrict a function to $A$.

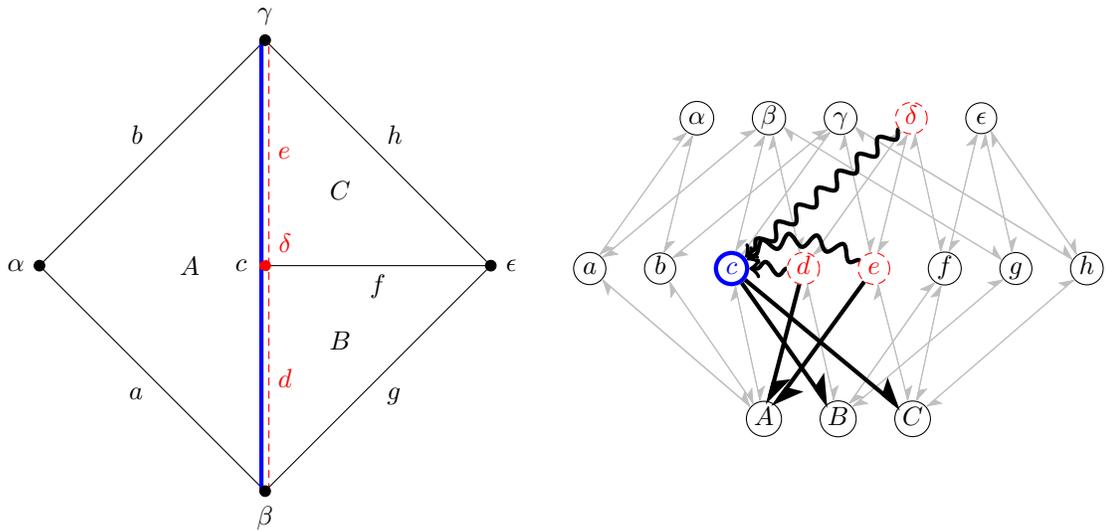
\begin{figure}\centering
  \tikzset{arrow/.append style={gray!50!white}}
  \tikzset{parent/.append style={blue,ultra thick}}
  \tikzset{child/.append style={red,densely dashed}}
  \tikzset{childlabel/.append style={red}}
  \begin{minipage}{0.49\textwidth}
    \begin{tikzpicture}
  \usetikzlibrary{thplex}
  \tikzset{parent/.prefix style={}}
  \tikzset{child/.prefix style={}}
  \tikzset{childlabel/.prefix style={}}
  \tikzset{parentlabel/.prefix style={}}
  \tikzset{star/.prefix style={}}
  \tikzset{starfill/.prefix style={}}

  \coordinate (al) at (-3, 0);
  \coordinate (be) at ( 0,-3);
  \coordinate (ga) at ( 0, 3);
  \coordinate (ep) at ( 3, 0);
  \coordinate (de) at ( 0, 0);
  \fill [cell,starfill] (al) -- (be) -- (ga) -- cycle
    node at (barycentric cs:al=1,be=1,ga=1) {$A$};
  \fill [cell] (be) -- (ep) -- (de) -- cycle
    node at (barycentric cs:be=1,ep=1,de=1) {$B$};
  \fill [cell,starfill] (ep) -- (ga) -- (de) -- cycle
    node at (barycentric cs:ep=1,ga=1,de=1) {$C$};
  \draw [facet] (al) -- node [below left] {$a$} (be);
  \draw [facet,star] (ga) -- node [above left] {$b$} (al);
  \draw [facet,shift left=0.5mm,parent] (be) -- node [facetlabel,parentlabel,left,xshift=-1mm] {$c$} (ga);
  \draw [facet,shift right=0.5mm,child] (be) -- node [facetlabel,childlabel,right,xshift=0.5mm] {$d$} (de);
  \draw [facet,shift right=0.5mm,star,child] (de) -- node [facetlabel,childlabel,right,xshift=0.5mm] {$e$} (ga);
  \draw [facet] (ep) -- node [below] {$f$} (de);
  \draw [facet] (be) -- node [below right] {$g$} (ep);
  \draw [facet,star] (ep) -- node [above right] {$h$} (ga);
  \node at (al) [vertex,label={[vertexlabel]left:$\alpha$}] {};
  \node at (be) [vertex,label={[vertexlabel]below:$\beta$}] {};
  \node at (ga) [vertex,star,label={[vertexlabel,star]above:$\gamma$}] {};
  \node at (de) [vertex,child,label={[vertexlabel,childlabel]above right:$\delta$}] {};
  \node at (ep) [vertex,label={[vertexlabel]right:$\epsilon$}] {};

\end{tikzpicture}

  \end{minipage}
  \begin{minipage}{0.49\textwidth}
    \tikzset{tree/.append style={ultra thick}}
    \begin{tikzpicture}[node distance=2cm]
  \usetikzlibrary{thplex}
  \tikzset{parent/.prefix style={}}
  \tikzset{child/.prefix style={}}
  \tikzset{newsupp/.prefix style={black,ultra thick}}

  \matrix (cells) {
    \node [point] (A) {$A$}; &
    \node [point] (B) {$B$}; &
    \node [point] (C) {$C$}; \\
  };
  \matrix [above of=cells] (facets) {
    \node [point] (a) {$a$}; &
    \node [point] (b) {$b$}; &
    \node [point,parent] (c) {$c$}; &
    \node [point,child] (d) {$d$}; &
    \node [point,child] (e) {$e$}; &
    \node [point] (f) {$f$}; &
    \node [point] (g) {$g$}; &
    \node [point] (h) {$h$}; \\
  };
  \matrix [above of=facets] (vertices) {
    \node [point] (al) {$\alpha$}; &
    \node [point] (be) {$\beta$};  &
    \node [point] (ga) {$\gamma$}; &
    \node [point,child] (de) {$\delta$}; &
    \node [point] (ep) {$\epsilon$}; \\
  };
  \draw[arrow] (A) -- (a);
  \draw[arrow] (a) -- (A);
  \draw[arrow] (A) -- (b);
  \draw[arrow] (b) -- (A);
  \draw[arrow] (A) -- (c);
  \draw[arrow] (c) -- (A);

  \draw[arrow] (B) -- (d);
  \draw[arrow] (d) -- (B);
  \draw[arrow] (B) -- (f);
  \draw[arrow] (f) -- (B);
  \draw[arrow] (B) -- (g);
  \draw[arrow] (g) -- (B);

  \draw[arrow] (C) -- (e);
  \draw[arrow] (e) -- (C);
  \draw[arrow] (C) -- (f);
  \draw[arrow] (f) -- (C);
  \draw[arrow] (C) -- (h);
  \draw[arrow] (h) -- (C);

  \draw[arrow] (a) -- (al);
  \draw[arrow] (al) -- (a);
  \draw[arrow] (a) -- (be);
  \draw[arrow] (be) -- (a);

  \draw[arrow] (b) -- (al);
  \draw[arrow] (al) -- (b);
  \draw[arrow] (b) -- (ga);
  \draw[arrow] (ga) -- (b);

  \draw[arrow] (c) -- (be);
  \draw[arrow] (be) -- (c);
  \draw[arrow] (c) -- (ga);
  \draw[arrow] (ga) -- (c);

  \draw[arrow] (d) -- (be);
  \draw[arrow] (be) -- (d);
  \draw[arrow] (d) -- (de);
  \draw[arrow] (de) -- (d);

  \draw[arrow] (e) -- (de);
  \draw[arrow] (de) -- (e);
  \draw[arrow] (e) -- (ga);
  \draw[arrow] (ga) -- (e);

  \draw[arrow] (f) -- (de);
  \draw[arrow] (de) -- (f);
  \draw[arrow] (f) -- (ep);
  \draw[arrow] (ep) -- (f);

  \draw[arrow] (g) -- (be);
  \draw[arrow] (be) -- (g);
  \draw[arrow] (g) -- (ep);
  \draw[arrow] (ep) -- (g);

  \draw[arrow] (h) -- (ga);
  \draw[arrow] (ga) -- (h);
  \draw[arrow] (h) -- (ep);
  \draw[arrow] (ep) -- (h);

  \draw[tree] (d) -- (c);
  \draw[tree] (e) to [bend right=30]  (c);
  \draw[tree] (de) -- (c);
  \draw[arrow,newsupp] (c) -- (B);
  \draw[arrow,newsupp] (c) -- (C);
  \draw[arrow,newsupp] (d) -- (A);
  \draw[arrow,newsupp] (e) -- (A);
\end{tikzpicture}

  \end{minipage}
  \caption{%
    An illustration of the extension of DMPlex for non-conformal meshes.
    The snaking lines show $\parent(d)$, $\parent(e)$, and $\parent(\delta)$;
    the bold support arrows do not have matching cone arrows, breaking the
    duality that is present in conformal meshes.%
  }%
  \label{fig:hasse}
\end{figure}

Given these considerations, we include both super-points and sub-points (or
``parents'' and ``children'' henceforth) in the representations of
non-conformal meshes in DMPlex.  In addition, we make the following extensions
to the format, illustrated in \cref{fig:hasse}:
\begin{enumerate}
  \item
    We add to the Hasse diagram, which is traversed with $\cone()$ and
    $\supp()$ operations, a separate tree structure, which is traversed with
    $\parent()$ and $\children()$ operations (in DMPlex, {\bf
    \pd{DM}{DMPlexGetTreeParent}()} and {\bf
    \pd{DM}{DMPlexGetTreeChildren}()}).
  \item
    We break the duality between $\cone()$ and $\supp()$ operations.  In
    particular, the cone of an $n$-cell $p$ includes the ``natural''
    decomposition of its boundary into $(n-1)$-cells (e.g., the three edges of
    a triangle), while the support of $p$ is the set of $(n+1)$-cells whose
    boundaries intersect $p$.  It is still the case that
    $q\in \cone(p)\Rightarrow p\in \supp(q)$, but the converse is not true for
    non-conformal meshes.  In the mesh in \cref{fig:hasse}, for example,
    $A\in \supp(d)$ because $\partial A  \cap d \neq \emptyset$, but
    $d\not\in\cone(A)$, because it is not one of the canonical edges of $A$.
    The extra support maps are included because:
    \begin{itemize}
      \item
        It ensures that $\Star(p)$ covers the support of $p$'s basis
        functions, which is important for determining the sparsity pattern of
        finite element matrices (if $\Star(p)\cap\Star(q)\neq\emptyset$, then
        there may be non-zeros entries in a finite element matrix for their
        degrees of freedom).
      \item
        It ensures that the support of a facet (a $(d-1)$-dimensional cell)
        can be used to identify neighboring cells for finite volume and
        discontinuous Galerkin methods.
    \end{itemize}
\end{enumerate}

These extensions pass the minimum bar of not affecting the behavior of DMPlex
for conformal meshes, but what we really want is for future extensions of
DMPlex designed for conformal meshes to work automatically for non-conformal
meshes as well.  Because the PETSc developers encourage contributions from
users, including to DMPlex (see \cite{LangeKnepleyGorman15} for a recent
example), this requires careful attention to the modifications to the DMPlex
interface, which we will discuss in the next section.

\subsection{The finite element method for non-conformal meshes}
\label{sec:nonconformalfem}

For a non-conformal mesh, a global nodal basis as defined in the previous
section is generally not possible: the union of all element functionals,
\begin{equation}\label{eq:unconstrained}
  W^u := \cup_{i=1}^{N_K} \cup_{p\in\hat{S}} \Sigma_i^p,
\end{equation}
will contain linear dependencies.  For a hierarchically non-conformal mesh
$S$, however, it is possible to construct a global basis $W^c$ that is nearly
nodal, by including only the functionals of points that have no ancestors,
\begin{equation}\label{eq:global}
  W^c := \cup_{i=1}^{N_K} \cup_{\textstyle
  \{p\in\hat{S}:\parent(\varphi_i(p))=\emptyset\}} \Sigma_i^p.
\end{equation}
There is then a \emph{constraint matrix} $I_c^u$ such that $W^u(v) = I_c^u
W^c(v)$ for all $v\in V_h$.   These are sometimes referred to as
``hanging-node'' constraints.  In this section, we describe the general method
for calculating $I_c^u$.

\begin{figure}\centering%
  \tikzset{>=latex}
  \begin{tikzpicture}
  \begin{scope}[shift={(-3,-0.5)},scale=1.5]
    \pgfsetfillcolor{gray!50!white}
    \pgfpathmoveto{\pgfpoint{-0.520cm}{0.854cm}}
    \pgfpatharcaxes{121.361}{180}{\pgfpoint{1.0cm}{0.0cm}}{\pgfpoint{0.0cm}{1.0cm}}
    \pgfpatharcaxes{15}{0}{\pgfpoint{-0.966cm}{-0.149cm}}{\pgfpoint{-0.259cm}{0.558cm}}
    \pgfpatharcaxes{0}{-35.264}{\pgfpoint{-0.966cm}{-0.149cm}}{\pgfpoint{0.000cm}{-0.816cm}}
    \pgfpatharcaxes{35.264}{0}{\pgfpoint{-0.612cm}{0.781cm}}{\pgfpoint{-0.5cm}{-0.5cm}}
    \pgfusepath{fill}
    \draw [dotted] (-0.520cm,0.854cm) arc[radius = 1,start angle=121.361,end angle=204.896];
    \pgfpathmoveto{\pgfpoint{-0.789cm}{0.349cm}}
    \pgfpatharcaxes{-35.264}{35.264}{\pgfpoint{-0.966cm}{-0.149cm}}{\pgfpoint{0.000cm}{-0.816cm}}
    \pgfpatharcaxes{-35.264}{35.264}{\pgfpoint{-0.612cm}{-0.373cm}}{\pgfpoint{0.5cm}{0.5cm}}
    \pgfpatharcaxes{-35.264}{35.264}{\pgfpoint{-0.259cm}{0.558cm}}{\pgfpoint{0.000cm}{0.816cm}}
    \pgfpatharcaxes{-35.264}{35.264}{\pgfpoint{-0.612cm}{0.781cm}}{\pgfpoint{-0.5cm}{-0.5cm}}
    \pgfpathmoveto{\pgfpoint{-1.577cm}{0.699cm}}
    \pgfpatharcaxes{-35.264}{35.264}{\pgfpoint{-1.932cm}{-0.299cm}}{\pgfpoint{0.000cm}{-1.633cm}}
    \pgfpathlineto{\pgfpoint{-0.789cm}{-0.593cm}}
    \pgfpathmoveto{\pgfpoint{-0.211cm}{0.927cm}}
    \pgfpathlineto{\pgfpoint{-0.423cm}{1.853cm}}
    \pgfpatharcaxes{-35.264}{35.264}{\pgfpoint{-1.225cm}{1.563cm}}{\pgfpoint{-1.0cm}{-1.0cm}}
    \pgfpathmoveto{\pgfpoint{-1.041cm}{1.708cm}}
    \pgfpatharcaxes{121.361}{204.896}{\pgfpoint{2.0cm}{0.0cm}}{\pgfpoint{0.0cm}{2.0cm}}
    \pgfpathmoveto{\pgfpoint{-0.966cm}{-0.149cm}}
    \pgfpathlineto{\pgfpoint{-0.483cm}{-0.075cm}}
    \pgfpatharcaxes{0}{35.264}{\pgfpoint{-0.483cm}{-0.075cm}}{\pgfpoint{0.0cm}{0.408cm}}
    \pgfpathlineto{\pgfpoint{-0.789cm}{0.349cm}}
    \pgfpathmoveto{\pgfpoint{-0.394cm}{0.175cm}}
    \pgfpatharcaxes{-35.264}{0}{\pgfpoint{-0.306cm}{0.391cm}}{\pgfpoint{0.25cm}{0.25cm}}
    \pgfpathlineto{\pgfpoint{-0.612cm}{0.781cm}}
    \pgfusepath{draw}
    \draw [dotted] (-0.789cm,0.349cm) -- (-1.577cm,0.699cm);
    \node (Ki) at (-0.526cm,0.473cm) {$K_i$};
    \node (Kj) at (-0.919cm,1.172cm) {$K_j$};
    \coordinate (C) at (-0.85cm,0.3cm);
  \end{scope}
  \begin{scope}[x = {(-150:1cm)},
                y = {(-30:1cm)},
                z = {(90:1cm)}]
    \tikzset{inject/.prefix style={fill=gray!50!white}}
    \fill [inject] (0,1,0) -- (1,1,0) -- (1,1,1) -- (0,1,1) -- cycle;
    \draw (1,-1,-1) -- (1,1,-1) -- (1,1,1) -- (1,-1,1) -- cycle;
    \draw (-1,1,-1) -- (1,1,-1) -- (1,1,1) -- (-1,1,1) -- cycle;
    \draw (-1,-1,1) -- (1,-1,1) -- (1,1,1) -- (-1,1,1) -- cycle;
    \coordinate (pre) at (1,-1,00);
    \coordinate (mid) at (0.5,1,0.5);
    \node (Kref) at (0,0,1) {$\hat{K}$};
  \end{scope}
  \draw [->] (Kref) to [bend right=30] node [above] {$\varphi_i$} (Ki);
  \draw [->] (Kref) to [bend right=30] node [above] {$\varphi_j$} (Kj);
  \draw [->] (pre) to [bend left=30] (C)
    to [bend right=39] node [below] {$\varphi_j^{-1} \circ \varphi_i$} (mid);
\end{tikzpicture}

  \caption{%
    Even though the embeddings $\varphi_i$ and $\varphi_j$ of $K_i$ and $K_j$
    are curvilinear, $\varphi_j^{-1} \circ \varphi_i$ is component-wise
    affine, so $H^1(\Omega)$-conforming spaces can be constructed
    from tensor-product polynomials on $\hat{K}$.%
  }%
  \label{fig:injection}
\end{figure}
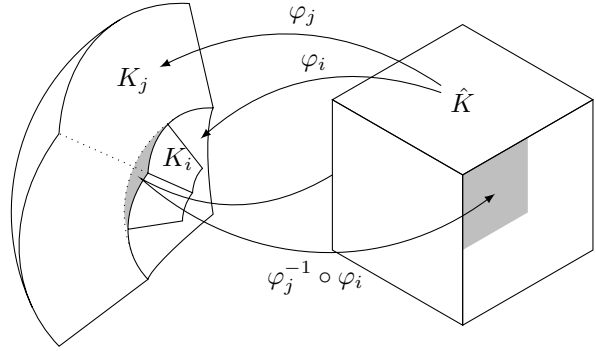

We retain assumptions I, II, and III from \cref{sec:conformal} when
considering non-conformal meshes.  Assumption II---that neighboring
approximations spaces ``line up'' for $H^1(\Omega)$-conforming
constructions---limits the types of non-conformal interfaces that can occur.
Given neighboring cells $K_i$ and $K_j$ and $p,q\in\hat{S}$ such that
$\varphi_i(p) \subset \varphi_j(q)$, then $\varphi_i^*\varphi_j^{-*}$ must map
$P(\varphi_j^{-1}\varphi_i(p))$ onto $P(p)$.  For simplicial elements with
polynomial spaces, this typically means $\varphi_j^{-1} \circ \varphi_i:p\to
q$ is affine; for hypercube elements with tensor-product polynomial spaces,
$\varphi_j^{-1} \circ \varphi_i$ must be component-wise affine.  We note that
this is not a requirement that $\varphi_i$ or $\varphi_j$ be affine
(\cref{fig:injection}).

When these conditions are met, we can expand each functional in $\Sigma_i^p$
in terms of functionals in $\Sigma_j$: for each $\sigma_r\in\hat{\Sigma}^p$,
\begin{align}
  (\varphi_{i*}\sigma_r)(v)
  &=
  (\varphi_{i*}\sigma_r)(\varphi_j^{-*}\varphi_j^*v)
  \\
  &=
  (\varphi_{j*}^{-1}\varphi_{i*}\sigma_r)(\varphi_j^*v)
  \\
  &=
  \sum_{\textstyle \sigma_s\in\hat{\Sigma}}
  (\varphi_{j*}^{-1}\varphi_{i*}\sigma_r)(\psi_s)\sigma_s(\varphi_j^*v)
  \\
  &=
  \sum_{\textstyle \sigma_s\in{\Sigma}_j}
  (\varphi_{j*}^{-1}\varphi_{i*}\sigma_r)(\psi_s)
  \sigma_s(v),\label{eq:functionalsum}
\end{align}
where $\psi_s$ is again the shape function of $\sigma_s$.  The transfered
functional $\varphi_{j*}^{-1}\varphi_{i*} \sigma_r$ can only be supported on
$q$, so $(\varphi_{j*}^{-1}\varphi_{i*} \sigma_r)(\psi_s)=0$ if
$q\not\in\supp(\psi_s)$.  By assumption I, this means that the terms in 
\cref{eq:functionalsum} are non-zero only if $\sigma_s\in\Sigma^t$ for
$t\in\clos(\parent(\varphi_i(p)))$,
\begin{align}\label{eq:pointconstraint}
  (\varphi_{i*}\sigma_r)(v)
  =
  \sum_{\textstyle \sigma_s\in\cup_{t\in\clos(\parent(\varphi_i(p)))}\Sigma^t}
  (\varphi_{j*}^{-1}\varphi_{i*}\sigma_r)(\psi_s)
  \sigma_s(v).
\end{align}
\Cref{eq:pointconstraint} illustrates the two key points needed to compute the
constraint matrix $I_c^u$:
\begin{enumerate}
  \item
    If $p\in S$ and $\parent(p)\neq \emptyset$, then functionals in $W^u$
    associated with $p$ are linear combinations of the functionals associated
    with points in $\clos(\parent(p))$.  If any of the points in that set has
    a parent, then we can iteratively apply $\clos\circ\parent$ to find $p$'s
    \emph{anchor points}, whose functionals will be in the global basis $W^c$.
    This lets us compute the sparsity pattern of $I_c^u$.
  \item
    The matrix that interpolates to $\Sigma^p$ from its anchor points'
    functionals has entries of the form $(\varphi_{j*}^{-1} \varphi_{i*}
    \sigma_r)(\psi_s)$ for $\sigma_r\in\hat{S}$ and shape function $\psi_s\in
    P(\hat{K})$.  This lets us compute entries in $I_c^u$.
\end{enumerate}

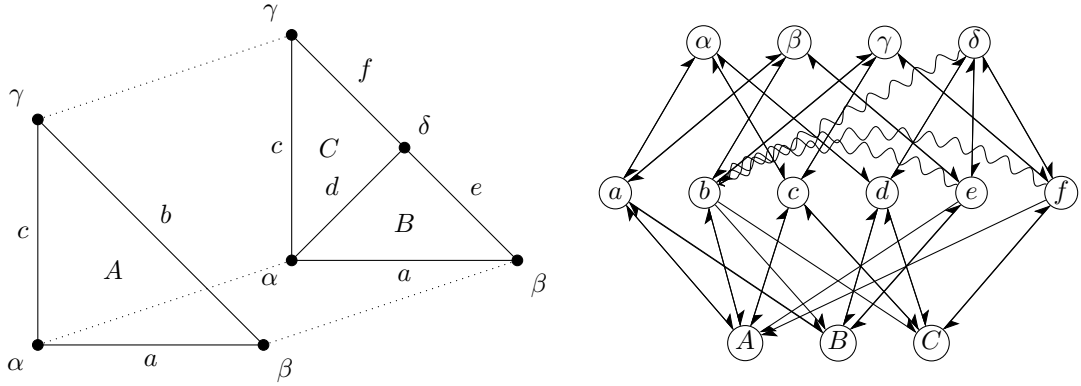
\begin{figure}
  \begin{minipage}{0.49\textwidth}\centering%
    \begin{tikzpicture}[scale=1.5]
  \usetikzlibrary{thplex}
  \begin{scope}[local bounding box=root]
    \coordinate (al) at (-1,-1); \node at (al) [vertex,label={[vertexlabel]below left:$\alpha$}] {};
    \coordinate (be) at ( 1,-1); \node at (be) [vertex,label={[vertexlabel]below right:$\beta$}] {};
    \coordinate (ga) at (-1, 1); \node at (ga) [vertex,label={[vertexlabel]above left:$\gamma$}] {};
    \fill [cell] (al) -- (be) -- (ga) -- cycle node at (barycentric cs:al=1,be=1,ga=1) {$A$};
    \draw [facet] (al) -- node [below]       {$a$} (be);
    \draw [facet] (be) -- node [above right] {$b$} (ga);
    \draw [facet] (ga) -- node [left]        {$c$} (al);
  \end{scope}
  \begin{scope}[shift={(2.25,0.75)},local bounding box=children]
    \coordinate (al2) at (-1,-1); \node at (al2) [vertex,label={[vertexlabel]below left:$\alpha$}] {};
    \coordinate (be2) at ( 1,-1); \node at (be2) [vertex,label={[vertexlabel]below right:$\beta$}] {};
    \coordinate (ga2) at (-1, 1); \node at (ga2) [vertex,label={[vertexlabel]above left:$\gamma$}] {};
    \coordinate (de) at  (0 , 0); \node at (de) [vertex,label={[vertexlabel]above right:$\delta$}] {};
    \fill [cell] (al2) -- (be2) -- (de) -- cycle node at (barycentric cs:al2=1,be2=1,de=1) {$B$};
    \fill [cell] (al2) -- (de) -- (ga2) -- cycle node at (barycentric cs:al2=1,de=1,ga2=1) {$C$};
    \draw [facet] (al2) -- node [below]       {$a$} (be2);
    \draw [facet] (be2) -- node [above right] {$e$} (de);
    \draw [facet] (de)  -- node [above right] {$f$} (ga2);
    \draw [facet] (ga2) -- node [left]        {$c$} (al2);
    \draw [facet] (al2) -- node [above left]  {$d$} (de);
  \end{scope}
  \draw[dotted] (al) -- (al2);
  \draw[dotted] (be) -- (be2);
  \draw[dotted] (ga) -- (ga2);
\end{tikzpicture}

  \end{minipage}
  \begin{minipage}{0.49\textwidth}\centering%
    \tikzset{every matrix/.append style={anchor=center,column sep=0.75cm}}
    \begin{tikzpicture}[node distance=2cm]
  \usetikzlibrary{thplex}
  \matrix (cells) {%
    \node [point] (hA) {$A$};
    &
    \node [point] (hB) {$B$};
    &
    \node [point] (hC) {$C$};
    \\
  };%
  \matrix (facets) [above of=cells]{%
    \node [point] (ha) {$a$};
    &
    \node [point] (hb) {$b$};
    &
    \node [point] (hc) {$c$};
    &
    \node [point] (hd) {$d$};
    &
    \node [point] (he) {$e$};
    &
    \node [point] (hf) {$f$};
    \\
  };%
  \matrix (vertices) [above of=facets]{%
    \node [point] (hal) {$\alpha$};
    &
    \node [point] (hbe) {$\beta$};
    &
    \node [point] (hga) {$\gamma$};
    &
    \node [point] (hde) {$\delta$};
    \\
  };%
  \draw[arrow] (hA) -- (ha);
  \draw[arrow] (ha) -- (hA);
  \draw[arrow] (hA) -- (hb);
  \draw[arrow] (hb) -- (hA);
  \draw[arrow] (hA) -- (hc);
  \draw[arrow] (hc) -- (hA);

  \draw[arrow] (hB) -- (ha);
  \draw[arrow] (ha) -- (hB);
  \draw[arrow] (hB) -- (hd);
  \draw[arrow] (hd) -- (hB);
  \draw[arrow] (hB) -- (he);
  \draw[arrow] (he) -- (hB);

  \draw[arrow] (hC) -- (hc);
  \draw[arrow] (hc) -- (hC);
  \draw[arrow] (hC) -- (hd);
  \draw[arrow] (hd) -- (hC);
  \draw[arrow] (hC) -- (hf);
  \draw[arrow] (hf) -- (hC);

  \draw[arrow] (ha) -- (hal);
  \draw[arrow] (hal) -- (ha);
  \draw[arrow] (ha) -- (hbe);
  \draw[arrow] (hbe) -- (ha);

  \draw[arrow] (hb) -- (hbe);
  \draw[arrow] (hbe) -- (hb);
  \draw[arrow] (hb) -- (hga);
  \draw[arrow] (hga) -- (hb);

  \draw[arrow] (hc) -- (hga);
  \draw[arrow] (hga) -- (hc);
  \draw[arrow] (hc) -- (hal);
  \draw[arrow] (hal) -- (hc);

  \draw[arrow] (hd) -- (hal);
  \draw[arrow] (hal) -- (hd);
  \draw[arrow] (hd) -- (hde);
  \draw[arrow] (hde) -- (hd);

  \draw[arrow] (he) -- (hbe);
  \draw[arrow] (hbe) -- (he);
  \draw[arrow] (he) -- (hde);
  \draw[arrow] (hde) -- (he);

  \draw[arrow] (hf) -- (hga);
  \draw[arrow] (hga) -- (hf);
  \draw[arrow] (hf) -- (hde);
  \draw[arrow] (hde) -- (hf);

  \draw[arrow] (hb) -- (hB);
  \draw[arrow] (hb) -- (hC);

  \draw[arrow] (he) -- (hA);
  \draw[arrow] (hf) -- (hA);

  \draw[tree,] (he) to [bend right=30] (hb);
  \draw[tree] (hf) to [bend right=30] (hb);
  \draw[tree] (hde) -- (hb);
\end{tikzpicture}
  \end{minipage}
  \caption{%
    The reference tree $\hat{T}$ for red-green refinement of triangles.
    Notice that the reference element $\hat{S}$ is included as a sub-complex
    of $\hat{T}$.%
  }%
  \label{fig:reftree}
\end{figure}

Without further information, entries in $I_c^u$ must be computed by evaluating
transfered functionals of the form $\varphi_{j*}^{-1}\varphi_{i*}\sigma$ for
any non-conformally adjacent cells $K_i$ and $K_j$.  In practice, however,
non-conformal meshes are usually generated by some predefined set of mesh
refinement rules.  These rules can be encoded in a small non-conformal mesh
that we call the \emph{reference tree} $\hat{T}$ .  For a mesh created by
red-green refinement (\cref{fig:reftree}), for example, every non-conformal
transfer map $\varphi_j^{-1} \circ \varphi_i$ is like mapping one of the edges
of the coarse cell ($a$, $b$, $c$) to one of the refined edges ($e$, $f$).
Due to symmetry, all we have to evaluate are the transfered functionals for
$(\varphi_j^{-1}\circ \varphi_i) \sim (b \mapsto e)$ and $(\varphi_j^{-1}
\circ \varphi_i) \sim (b \mapsto f)$, which can then be copied into the
correct locations in $I_c^u$.

\section{The DMPlexTree interface}

Support for non-conformal meshes in DMPlex is available in the latest release
of PETSc (v3.6), and full documentation can be found online.  In this section
we introduce the most important components of the interface, starting with the
highest-level methods that require the least intervention from the user, and
descending into some of the finer controls available to experts.

\begin{figure}\centering
  \begin{minipage}{0.24\textwidth}\centering
    \begin{tikzpicture}
      \usetikzlibrary{thplex}
      \coordinate (a2) at (-1,-1); \node at (a2) [vertex] {};
      \coordinate (b2) at ( 1,-1); \node at (b2) [vertex] {};
      \coordinate (c2) at (-1, 1); \node at (c2) [vertex] {};
      \coordinate (d)  at ( 0,-1); \node at (d)  [vertex] {};
      \coordinate (e)  at ( 0, 0); \node at (e)  [vertex] {};
      \coordinate (f)  at (-1, 0); \node at (f)  [vertex] {};
      \fill [cell]  (a2) -- (d) -- (f) -- cycle;
      \fill [cell]  (b2) -- (e) -- (d) -- cycle;
      \fill [cell]  (c2) -- (f) -- (e) -- cycle;
      \fill [cell]  (d)  -- (e) -- (f) -- cycle;
      \draw [facet] (a2) -- (d) -- (b2) -- (e) -- (c2) -- (f) -- cycle;
      \draw [facet] (d) -- (e) -- (f) -- cycle;
    \end{tikzpicture}
  \end{minipage}%
  \begin{minipage}{0.24\textwidth}\centering
    \begin{tikzpicture}
      \usetikzlibrary{thplex}
      \coordinate (a2) at (-1,-1); \node at (a2) [vertex] {};
      \coordinate (b2) at ( 1,-1); \node at (b2) [vertex] {};
      \coordinate (c2) at ( 1, 1); \node at (c2) [vertex] {};
      \coordinate (d2) at (-1, 1); \node at (d2) [vertex] {};
      \coordinate (e)  at ( 0,-1); \node at (e)  [vertex] {};
      \coordinate (f)  at ( 1, 0); \node at (f)  [vertex] {};
      \coordinate (g)  at ( 0, 1); \node at (g)  [vertex] {};
      \coordinate (h)  at (-1, 0); \node at (h)  [vertex] {};
      \coordinate (i)  at ( 0, 0); \node at (i)  [vertex] {};
      \fill [cell]  (a2) -- (e) -- (i) -- (h) -- cycle;
      \fill [cell]  (b2) -- (f) -- (i) -- (e) -- cycle;
      \fill [cell]  (c2) -- (g) -- (i) -- (f) -- cycle;
      \fill [cell]  (d2) -- (h) -- (i) -- (g) -- cycle;
      \draw [facet] (a2) -- (e) -- (b2) -- (f) -- (c2) -- (g) -- (d2)
                         -- (h) -- cycle;
      \draw [facet] (e) -- (i) -- (g);
      \draw [facet] (f) -- (i) -- (h);
    \end{tikzpicture}
  \end{minipage}%
  \begin{minipage}{0.24\textwidth}\centering
    \begin{tikzpicture}[x={(-65.105:0.753cm)},y={(8.794:1.197cm)},z={(90:1cm)}]
      \coordinate (a2) at (-1,-1,-1); \node at (a2) [vertex] {};
      \coordinate (b2) at ( 1,-1,-1); \node at (b2) [vertex] {};
      \coordinate (c2) at (-1, 1,-1); \node at (c2) [vertex] {};
      \coordinate (d2) at (-1,-1, 1); \node at (d2) [vertex] {};
      \coordinate (e)  at ( 0,-1,-1); \node at (e)  [vertex] {};
      \coordinate (f)  at ( 0, 0,-1); \node at (f)  [vertex] {};
      \coordinate (g)  at (-1,-1, 0); \node at (g)  [vertex] {};
      \coordinate (h)  at ( 0,-1, 0); \node at (h)  [vertex] {};
      \coordinate (i)  at (-1, 0, 0); \node at (i)  [vertex] {};
      \fill[cell] (a2) -- (e) -- (g) -- cycle;
      \fill[cell] (b2) -- (h) -- (e) -- cycle;
      \fill[cell] (d2) -- (g) -- (h) -- cycle;

      \fill[cell] (b2) -- (f) -- (h) -- cycle;
      \fill[cell] (c2) -- (i) -- (f) -- cycle;
      \fill[cell] (d2) -- (h) -- (i) -- cycle;

      \draw[facet] (a2) -- (b2) -- (d2) -- cycle;
      \draw[facet] (b2) -- (c2) -- (d2) -- cycle;
      \draw[facet] (e) -- (g) -- (h) -- cycle;
      \draw[facet] (h) -- (i) -- (f) -- cycle;
    \end{tikzpicture}
  \end{minipage}%
  \begin{minipage}{0.24\textwidth}\centering
    \begin{tikzpicture}[x={(-65.105:0.753cm)},y={(8.794:1.197cm)},z={(90:1cm)}]
      \coordinate (a) at (-1,-1, 1); \node at (a) [vertex] {};
      \coordinate (b) at ( 0,-1, 1); \node at (b) [vertex] {};
      \coordinate (c) at ( 1,-1, 1); \node at (c) [vertex] {};
      \coordinate (d) at (-1, 0, 1); \node at (d) [vertex] {};
      \coordinate (e) at ( 0, 0, 1); \node at (e) [vertex] {};
      \coordinate (f) at ( 1, 0, 1); \node at (f) [vertex] {};
      \coordinate (g) at (-1, 1, 1); \node at (g) [vertex] {};
      \coordinate (h) at ( 0, 1, 1); \node at (h) [vertex] {};
      \coordinate (i) at ( 1, 1, 1); \node at (i) [vertex] {};
      \coordinate (j) at (-1,-1, 0); \node at (j) [vertex] {};
      \coordinate (k) at ( 0,-1, 0); \node at (k) [vertex] {};
      \coordinate (l) at ( 1,-1, 0); \node at (l) [vertex] {};
      \coordinate (m) at ( 1, 0, 0); \node at (m) [vertex] {};
      \coordinate (n) at ( 1, 1, 0); \node at (n) [vertex] {};
      \coordinate (o) at (-1,-1,-1); \node at (o) [vertex] {};
      \coordinate (p) at ( 0,-1,-1); \node at (p) [vertex] {};
      \coordinate (q) at ( 1,-1,-1); \node at (q) [vertex] {};
      \coordinate (r) at ( 1, 0,-1); \node at (r) [vertex] {};
      \coordinate (s) at ( 1, 1,-1); \node at (s) [vertex] {};
      \fill[cell] (a) -- (b) -- (e) -- (d) -- cycle;
      \fill[cell] (b) -- (c) -- (f) -- (e) -- cycle;
      \fill[cell] (e) -- (f) -- (i) -- (h) -- cycle;
      \fill[cell] (d) -- (e) -- (h) -- (g) -- cycle;

      \fill[cell] (a) -- (b) -- (k) -- (j) -- cycle;
      \fill[cell] (b) -- (c) -- (l) -- (k) -- cycle;
      \fill[cell] (k) -- (l) -- (q) -- (p) -- cycle;
      \fill[cell] (j) -- (k) -- (p) -- (o) -- cycle;

      \fill[cell] (c) -- (f) -- (m) -- (l) -- cycle;
      \fill[cell] (f) -- (i) -- (n) -- (m) -- cycle;
      \fill[cell] (m) -- (n) -- (s) -- (r) -- cycle;
      \fill[cell] (l) -- (m) -- (r) -- (q) -- cycle;

      \draw[facet] (a) -- (b) -- (c) -- (f) -- (i) --
                   (i) -- (h) -- (g) -- (d) -- (a) --
                   (j) -- (o) -- (p) -- (q) -- (r) --
                   (s) -- (n) -- (i)
                   (h) -- (e) -- (b) -- (k) -- (p)
                   (j) -- (k) -- (l) -- (m) -- (n)
                   (d) -- (e) -- (f) -- (m) -- (r)
                   (c) -- (l) -- (q);
    \end{tikzpicture}
  \end{minipage}%

  \caption{%
    The refinement patterns for the reference trees created by {\bf
    \protect\pd{DM}{DMPlexCreateDefaultReferenceTree}()}.  The reference trees
    themselves also contain the original coarse cells.
  }%
  \label{fig:defaulttrees}
\end{figure}
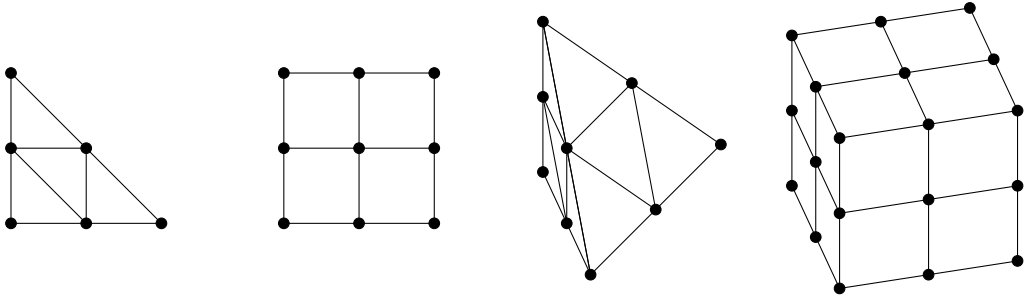

In \cref{sec:nonconformalfem}, we described how the existence of a predefined
refinement pattern, encoded in a reference tree $\hat{T}$, can enable DMPlex
to compute the constraint matrix $I_c^u$ more efficiently.  The reference tree
is also represent by a DMPlex that is assigned to the target mesh
(\cref{alg:setreftree}).  Reference tree implementations for isotropic
refinement on simplices and hypercubes for $d=1,2,3$ are provided by PETSc
(\cref{fig:defaulttrees}).

\begin{algorithm}
  \hypersetup{urlbordercolor={1 1 1}}
  \caption{%
    Creating a simplicial mesh {\tt dm} and setting the reference tree%
  }%
  \label{alg:setreftree}

  MPI\_Comm \ \ \ \ comm = PETSC\_COMM\_WORLD;

  \pd{Sys}{PetscBool} \ \ \ isSimplicial = PETSC\_TRUE;

  \pd{DM}{DM} \ \ \ \ \ \ \ \ \ \ dm, refTree;

  \pd{IS}{PetscSection} parentSection;

  \pd{Sys}{PetscInt} \ \ \ \ *parents, *childIDs;

  \null

  \tcc{Create a DMPlex, using, e.g., \pd{DM}{DMPlexCreateFromDAG}()}
  \ierr \fn{DMPlexCreate\dots}(comm,\dots,\&dm);\chkerr

  \tcc{Create a reference tree that describes the type of non-conformal
  interfaces in the mesh}

  \ierr \fn{\pd{DM}{DMGetDimension}}(dm,\&dim);\chkerr

  \ierr \fn{\pd{DM}{DMPlexCreateDefaultReferenceTree}}(comm,dim,isSimplicial,\&refTree);
  \chkerr

  \ierr \fn{\pd{DM}{DMPlexSetReferenceTree}}(dm,refTree);\chkerr

  \tcc{dm retains a reference to refTree, this reference can be destroyed}
  \ierr \fn{\pd{DM}{DMDestroy}}(\&refTree);\chkerr
\end{algorithm}

One can create a conformal DMPlex mesh from just the cone maps ({\bf
\pd{DM}{DMPlexCreateFromDAG}()}), and DMPlex can infer the support maps:
likewise, one can create a non-conformal mesh from just the parent maps.  In
\cref{alg:settree}, we demonstrate setting up the parent maps for the simple
non-conformal mesh in \cref{fig:mesh}, assuming that the red-green refinement
in \cref{fig:reftree} is used as a reference tree.

\begin{algorithm}
  \hypersetup{urlbordercolor={1 1 1}}
  \caption{%
    Setting the parent() maps and child IDs for {\tt dm}
  }%
  \label{alg:settree}
  \tcc{The figures use symbols for points, but we have to assign numbers to
  them.  We count across each stratum, starting at the bottom.}
  \pd{Sys}{PetscInt} \ \ \ \ numPoints = 16, c = 5, d = 6, e = 7, delta = 14;

  \tcc{c is the parent of each of the children}
  \pd{Sys}{PetscInt} \ \ \ \ parents[3] = \{c, c, c\};

  \tcc{Set numbers for the relevant points in the reference tree as well.}
  \pd{Sys}{PetscInt} \ \ \ \ bRef = 4, eRef = 7, fRef = 8, deltaRef = 12;

  \tcc{the childIDs are the points in the reference tree to which the children
  are analogous. d is to its parent (c) as eRef is to its parent (bRef), so
  that is its childID.}

  \pd{Sys}{PetscInt} \ \ \ \ childIDs[3] = \{eRef, fRef, deltaRef\};

  \pd{IS}{PetscSection} pSec;

  MPI\_Comm \ \ \ \ comm = \pd{Sys}{PetscObjectComm}((\pd{Sys}{PetscObject})dm);

  \null

  \ierr \fn{\pd{IS}{PetscSectionCreate}}(comm,\&pSec);\chkerr

  \ierr \fn{\pd{IS}{PetscSectionSetChart}}(pSec,0,numPoints);\chkerr

  \ierr \fn{\pd{IS}{PetscSectionSetDof}}(pSec,d,1);\chkerr

  \ierr \fn{\pd{IS}{PetscSectionSetDof}}(pSec,e,1);\chkerr

  \ierr \fn{\pd{IS}{PetscSectionSetDof}}(pSec,delta,1);\chkerr

  \ierr \fn{\pd{IS}{PetscSectionSetUp}}(pSec);\chkerr

  \ierr \fn{\pd{DM}{DMPlexSetTree}}(dm,pSec,parents,childIDs);\chkerr

  \ierr \fn{\pd{DM}{PetscSectionDestroy}}(\&pSec);\chkerr
\end{algorithm}

A DM can encompass not only a mesh, but also the fields discretized on it,
using a common interface for both finite element and finite volume methods.
In \cref{alg:setfem}, we set a standard Lagrange $\mathbb{P}_1(\hat{K})$
finite element on a mesh.

\begin{algorithm}
  \hypersetup{urlbordercolor={1 1 1}}
  \caption{Adding a finite element to a mesh {\tt dm}}
  \label{alg:setfem}
  \tcc{We are creating a scalar field}
  \pd{Sys}{PetscInt} \ \ \ numComp = 1;

  \tcc{We have a simplicial reference element}
  \pd{Sys}{PetscBool} \ \ isSimplex = PETSC\_TRUE;

  \tcc{The options prefix, for setting options at runtime: e.g., one can
  change the approximation order with `-my\_fe\_petscspace\_order 2`}

  const char *prefix = "my\_fe\_";

  \tcc{The quadrature order}
  \pd{Sys}{PetscInt} \ \ \ qorder = 1;

  \pd{Sys}{PetscInt} \ \ \ dim;

  \pd{DM}{PetscFE} \ \ \ \ fe;
  
  \null

  \ierr \fn{\pd{DM}{DMGetDimension}}(dm,\&dim);\chkerr

  \ierr \fn{\pd{DM}{PetscFECreateDefault}}(dm,dim,numComp,isSimplex,prefix,qorder,\&fe);\chkerr

  \ierr \fn{\pd{DM}{DMSetField}}(dm,0,(PetscObject) fe);\chkerr

  \ierr \fn{\pd{DM}{PetscFEDestroy}}(\&fe);\chkerr
\end{algorithm}

With the reference tree ({\bf \pd{DM}{DMPlexSetReferenceTree}()}), the parent
maps ({\bf \pd{DM}{DMPlexSetTree}()}), and the finite element ({\bf
\pd{DM}{DMSetField}()}), PETSc will:
\begin{itemize}
  \item
    determine the size of the global vector space (the size of $W^c$ in
    \cref{eq:global}),
  \item
    compute the constraint matrix $I_c^u$ from point constraints
    (\cref{eq:pointconstraint}),
  \item
    apply $I_c^u$ when getting the local form of a vector
    (\cref{alg:feresidual}, line \ref{line:globaltolocal}), so that the local
    form represent the vector evaluated at the unconstrained functionals $W^u$
    (\cref{eq:unconstrained}), and DMPlexVecGetClosure() gets the vector
    evaluated at the element functionals $\Sigma_i$ (\cref{eq:pushforward}),
  \item
    apply $I_c^{u\tr}$ when combining local residuals into a global
    residual (\cref{alg:feresidual}, line \ref{line:localtoglobal}),
  \item
    transform element matrices by the constraints in {\bf
    \pd{DM}{DMPlexMatSetClosure}()} to correctly assemble a global
    system matrix from element matrices.
\end{itemize}

To create a non-conformal mesh that uses a different refinement pattern than
the ones provided by PETSc, the user can create a custom reference tree.  Any
DMPlex that has had the parent maps set with {\bf \pd{DM}{DMPlexSetTree}()}
can serve as a reference tree.

If the user does not provide a finite element, then DMPlex cannot determine
for itself the layout of the vector space, and the entries in the constraint
matrix $I_c^u$ cannot be calculated automatically.  If the user specifies the
number of degrees of freedom associated with each process-local mesh point
(using {\bf \pd{DM}{DMSetDefaultSection}()}) then the tree data can be
used to both compute the size of the global vector space and the sparsity
pattern of $I_c^u$.  The user can then fill the entries of $I_c^u$ manually.

There are also potential uses for intra-mesh constraints between degrees of
freedom that do not fit into the hierarchical non-conformal framework that is
our focus here.  The constraint matrix $I_c^u$ can be a PETSc Mat of any
specification, and can be added to a DM directly with {\bf
\pd{DM}{DMSetDefaultConstraints}()}.  These constraints are applied at the
conclusion of {\bf \pd{DM}{DMGlobalToLocalEnd}()}, which gets the
process-local representation of a vector, and the transpose of these
constraints are applied at the beginning of {\bf
\pd{DM}{DMLocalToGlobalBegin}()}, when the contributions of all processes are
summed into a global vector.

\section{Verification and Example Usage}

In DMPlex's example program ex3,%
\footnote{%
  {\tt src/dm/impls/plex/examples/tests/ex3.c}: run {\tt make ex3} in that
  directory to build the example.
}
we include a small verification that DMPlex handles non-conformal meshes
properly.  The example can be run to create simplicial or hypercube meshes
with non-conformal interfaces.%
\footnote{%
  The initial intent of our work is merely to allow non-conformal meshes to be
  represented in DMPlex, not to implement a stand-alone adaptive mesh
  refinement interface.  To test the DMPlexTree interface without relying on
  external libraries, however, we have written {\bf
  \pd{DM}{DMPlexTreeRefineCell}()},
  which hierarchically refines a single cell of a conformal mesh.%
}%
\footnote{%
  To visualize the non-conformal meshes used, go to the {\tt
  src/dm/impls/plex/examples/tests/} directory of the PETSc source and run
  {\tt make ex3; ./ex3 -tree -simplex B -dim D -dm\_view
  vtk:nonconf\_B\_D.vtk:ASCII\_VTK}
  for ${\tt B}\in\{{\tt0},{\tt1}\}$ and ${\tt D}\in\{{\tt 2},{\tt 3}\}$.
}
To test that the finite element computations are handled correctly, we
construct a symmetric-gradient Laplacian operator $E$,
\begin{equation}\label{eq:symgrad}
  E(u,v) := \int_{\Omega} \half(\nabla u + \nabla u^{\tr}) :
  \half(\nabla v + \nabla v^{\tr})\ dx,
\end{equation}
and check whether rigid-body motions are in the null-space of $E$.  The
rigid-body motions will be in the null-space of each element matrix that is
computed, but if $I_c^u$ is incorrect, then they will be summed into the
global system matrix incorrectly, and it is very unlikely that they will be in
be in the null-space of the incorrect matrix (\cref{alg:ex3}).
\begin{algorithm}
  \hypersetup{urlbordercolor={1 1 1}}
  \caption{%
    Testing the correctness of an assembled Jacobian for a non-conformal
    mesh {\tt dm} (abridged from ex3.c)%
  }%
  \label{alg:ex3}
  \pd{Mat}{MatNullSpace} sp;

  \pd{Vec}{Vec} \ \ \ \ \ \ \ \ \ local;

  \pd{Sys}{PetscBool} \ \ \ isNullSpace;

  \null

  \tcc{This tests that the global system size is determined correctly, and
  that the sparsity pattern for global system matrices is computed correctly}
  \ierr \fn{\pd{DM}{DMCreateMatrix}}(dm,\&E);\chkerr

  \ierr \fn{\pd{DM}{DMGetLocalVector}}(dm,\&local);\chkerr

  \tcc{This is a finite-element loop within PETSc's SNES library that
  assembles the Jacobian matrices of nonlinear equations: the vector local is
  needed as a dummy argument to represent the current "solution" used to
  evaluate the Jacobian, which in this case is the linear operator in
  \cref{eq:symgrad}.  The variational form needed to compute each element's
  matrix has already been attached to dm.}
  \ierr \fn{\pd{SNES}{DMPlexSNESComputeJacobianFEM}}(dm,local,E,E,\nll);\chkerr

  \ierr \fn{\pd{DM}{DMPlexCreateRigidBody}}(dm,\&sp);\chkerr

  \ierr \fn{\pd{Mat}{MatNullSpaceTest}}(sp,E,\&isNullSpace);\chkerr
\end{algorithm}

\begin{figure}\centering
  \includegraphics[width=0.5\textwidth]{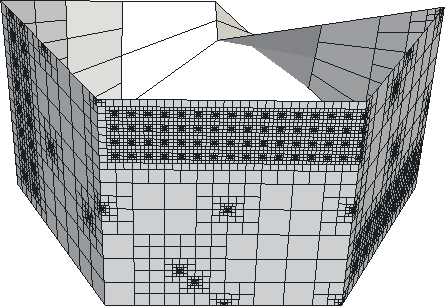}
  \caption{%
    A M\"{o}bius strip mesh, generated in p4est and converted to DMPlex in the
    example program {\tt p4est\_test\_plex}.
  }%
  \label{fig:p4est}
\end{figure}

Example usage outside of PETSc can be found in the p4est library for parallel
adaptive mesh refinement \cite{burstedde2011p4est}, which implements the
forest-of-quadtrees and forest-of-octrees paradigms in 2D and 3D.  This
library is meant to provide data structures only, and comes with no built-in
solver or finite element framework.  By converting the p4est format into
DMPlex (building on the methods described in \cite{IsaacBursteddeWilcoxEtAl15}
for efficiently converting p4est's native format to adjacency-based formats
like DMPlex), we make PETSc's numerical methods more readily available to
p4est users.  Example programs that perform this conversion are distributed
with p4est as {\tt p4est\_test\_plex} (2D) (\cref{fig:p4est}) and {\tt
p8est\_test\_plex} (3D).  The repository of the p4est
library\footnote{\url{https://bitbucket.org/cburstedde/p4est/}} has a
``petsc'' branch that is compatible with PETSc 3.6.
\footnote{%
  To build these examples, run {\tt ./configure --with-petsc=\$PETSC\_DIR}
  and {\tt make test/p4est\_test\_plex test/p8est\_test\_plex}.  To view the
  DMPlex meshes created in these tests, run the examples with the flag
  {\tt -dm\_view vtk:p4est\_petsc.vtk:ASCII\_VTK}.%
}

\section{Other discretizations}

The focus of this work has been $H^1(\Omega)$-conforming finite elements.  We
briefly discuss the way our approach to non-conformal meshes in DMPlex affects
other discretizations.

\subsection{$H^{\Curl}(\Omega)$-{} and $H^{\Div}(\Omega)$-conforming finite
elements}
\label{sec:curldiv}

PetscFE does not currently implement the covariant and contravariant Piola
transforms that are commonly used by $H^{\Curl}(\Omega)$-{} and
$H^{\Div}(\Omega)$-conforming finite elements, but these methods can still be
formulated via pullback onto reference elements \cite{rognes2009efficient}, so
future PetscFE implementations of these finite elements are a possibility.  The
discussion of conformal meshes and non-conformal meshes in this work is still
valid for these finite elements, with two small modifications:
\begin{itemize}
  \item
    The pullback operations are defined to be
    \begin{align}
      \varphi_i^* v
      &:=
      \Nabla \varphi_i^{\tr} v \circ \varphi_i
      &[\text{covariant, }H^{\Curl}(\Omega)],
      \\
      \varphi_i^* v
      &:=
      |\text{det} \Nabla \varphi_i| \Nabla \varphi_i v \circ \varphi_i
      &[\text{contravariant, }H^{\Div}(\Omega)].
    \end{align}
  \item
    The trace space $P(p)$ for a point $p\in\hat{S}$ involves not only
    restricting the function space $P(\hat{K})$ to the point, but also
    restricting to the tangential component ($H^{\Curl}$) or the normal
    component ($H^{\Div}$).
\end{itemize}

The core operation to compute continuity constraints is the transfer of a
functional from one element $K_i$ to its neighbor $K_j$ and evaluation on a
shape function, $(\varphi_{j*}^{-1} \varphi_{i*}\sigma)(\psi) =
\sigma(\varphi_i^*\varphi_j^{-*}\psi)$.  After one verifies that
$\varphi_i^*\varphi_j^{-*}=(\varphi_j^{-1} \circ \varphi_i)^*$ for both
pullbacks above, then it must be true that if each child-to-parent map
$\varphi_j^{-1} \circ \varphi_i$ is represented in the reference tree
$\hat{T}$, then it can be used to compute the entries in the constraint matrix
$I_c^u$.

\subsection{The finite volume method}

Finite volume methods do not promote the encapsulation of complexity as well
as finite element methods.  We have formulated our non-conformal mesh
extension for finite elements such that, in a typical finite element loop
(\cref{alg:feresidual}), the operations performed on each cell in the loop do
not depend on whether or not any of the points in the cell's closure is a
child or a parent.  In a cell-centric approach to the finite volume method,
the act of reconstructing centroid values requires determining the neighbors
of a cell, which becomes more complex when multiple cells may be on the
opposite side of a face (in \cref{fig:mesh}, e.g., both cells $B$ and $C$ are
opposite cell $A$ across edge $c$).  While we are currently incorporating
non-conformal meshes into the finite volume method as implemented by PetscFV
(we expect to finish while this manuscript is in review), the result is likely
to be more fragile to user extensions. 

One particular aspect that will be counterintuitive to users who worked with
finite volume methods on conformal meshes is that a facet can have more than
two cells in its support.  One often finds in finite volume code constructs of
the form ``{\tt neighbor = (supp[0] == me) ? supp[1] : supp[0]},'' which are
no longer valid.  One also has to avoid double-counting fluxes, i.e.,
computing fluxes on both a parent facet and its children. 

Because many unstructured finite volume methods do not care about the shape of
cells (i.e., whether they are triangles or quadrilaterals), these issues can
be avoided by encoding non-conformal meshes as conformal (though degenerate)
ones, as in \cref{fig:hassebad}.  In a multiphysics setting, where a finite
volume field and a finite element field are involved in a larger system of
equations, this approach is not possible.

\section{Discussion}

We have presented an extension to PETSc's DMPlex interface for unstructured
meshes so that it can now represent hierarchical non-conformal meshes.  Our
extension leaves the interface for conformal meshes the same, but adds a tree
structure to encode the hierarchy of subsets (children) and supersets
(parents).  We have shown how, for a wide class of finite elements, by
combining this hierarchical information with a reference tree that describes
the types of non-conformal interfaces that appear in a mesh, the extra
complexity of non-conformal meshes can be hidden, allowing finite element code
written for conformal meshes to be applied to them.  This extension can
already be used to convert p4est forest-of-quadtrees and forest-of-octrees
meshes to DMPlex.  Work is underway to bring support for the finite volume
method up to the level of the finite element method, and future work on the
discontinuous Galerkin method is planned.

\section{Acknowledgments}

We gratefully acknowledge the support of the Intel Parallel Computing Center
at the University of Chicago.

\bibliographystyle{ACM-Reference-Format-Journals}
\bibliography{thplex}

\end{document}